\documentclass[aps,pra,showpacs,twoside,twocolumn,longbibliography,10pt]{revtex4-1}
\usepackage[colorlinks=true, citecolor=red, urlcolor=blue ]{hyperref}
\usepackage{braket}
\usepackage{color}
\usepackage{graphicx}
\usepackage{dcolumn}
\usepackage{bbm}
\usepackage{caption}
\usepackage{subcaption}

\usepackage{amsmath}
\usepackage{amssymb}

\def\be{\begin{equation}}
\def\ee{\end{equation}}
\def\bea{\begin{eqnarray}}
\def\eea{\end{eqnarray}}
\DeclareMathOperator{\Tr}{Tr}
\usepackage[justification=justified,format=plain,singlelinecheck=true]{caption}

\begin{document}

\title{Disappearance of measurement-induced phase transition in a quantum spin system for large sizes}

\author{Paranjoy Chaki$^1$, Protyush Nandi$^2$, Ujjwal Sen$^1$ and Subinay Dasgupta$^1$}

\affiliation{$^1$Harish-Chandra Research Institute,  A CI of Homi Bhabha National Institute, Chhatnag Road, Jhunsi, Prayagraj 211 019, India,\\ $^2$University of Calcutta, 92, Acharya Prafulla Chandra Road, Kolkata - 700009}

\begin{abstract}
         Quantum measurements cause an abrupt change, called measurement induced phase transition, in the scaling behaviour of entanglement entropy in a quantum many-body system. The phenomenon is often studied in random quantum circuits, with local measurements performed with a certain probability. We report here one such  transition in an interacting spin 1/2 system where a global measurement is performed with certainty at every time-step. We start with a pure state with all the spins up. Each time step consists of evolution under the transverse Ising Hamiltonian for a time $\tau$, followed by a measurement that provides a ``yes/no'' answer to the question, ``Are all spins up?''. For various $\tau$ values, we compute the survival probability of the inital state, entanglement in bipartition, and the generalized geometric measure (a genuine multiparty entanglement), for a chain of size $L \sim 28$, and identify the transition points $\tau_c$ for  different field strengths. We then analytically derive a recursion relation that enables us to calculate the survival probability for system sizes up to 1000, and provides evidence of a scaling $\tau_c \sim 1/\sqrt{L}$. Therefore, the transition at finite $\tau_c$ for $L \sim 28$ seems to recede to $\tau_c = 0$ in the thermodynamic limit. Our work prompts a widespread investigation on the existence of measurement-induced transitions at large system-size. 
\end{abstract}
\maketitle

{ Measurement hypothesis of Quantum Mechanics indicates that  measurement of a variable on a quantum state causes the state to collapse to an eigenstate of the measurement operator. When this change is studied as a function of the control parameter, sometimes it so happens that the change is qualitatively different on two sides of some special values of the parameters. Such phenomena are called measurement-induced phase transition, in analogy with the change in physical properties of a thermodynamic system at a phase transition. A paradigmatic model for studying measurement-induced phase transition (MIPT) is a quantum gate circuit of some suitable structure in which measurements are performed locally \cite{zeno_1,Skinner,MIPT_2,MIPT_3,Chan2019,Szyniszewski2019,Bao2020,Choi2020,Gullans2020,GullansHuse2020,Jian2020,Zabalo2020,Iaconis2020,Turkeshi2020,Zhang2020,SzyniszewskiII2020,Nahum2021,Ippoliti2021,IppolitiII2021,lavasani2021,LavasaniII2021,Sang2021,Block2022,Sharma2022,Agrawal2022,Baratt2022,jian2023,Shane2023}, (for review, see \cite{fisher2023}). The time evolution of the system consists of unitary evolution under some operator $U$ for an interval of time, followed by (projective or weak) measurement of some suitable local operator with probability ${\mathcal P}$. After every time step, the bipartite entanglement entropy $S(t)$ is measured. 
It is found that in the asymptotic limit of large size and long time, for small values of ${\mathcal P}$, $S(t)$ increases linearly with time (``volume law''), while for large values of ${\mathcal P}$ it saturates after some time (``area law''). At some intermediate value ${\mathcal P}={\mathcal P}_c$, the entropy increases logarithmically with time, and this is considered to be a signature of a phase transition that has been induced by the process of measurement. The occurrence of such transitions has been confirmed in numerous systems in the last five years, and most of them are `small' in size ($<40$) \cite{Skinner,Chan2019,Szyniszewski2019,Bao2020,Choi2020,U_mipt}, while some are larger ($> 100$) \cite{zeno_1,MIPT_3,Gullans2020,GullansHuse2020}. The location of the transition point ${\mathcal P}_c$ is believed to be independent of size. 
{Apart from the random circuit analysis, this type of transition is also captured in the transverse field Ising model ~\cite{Isi_1,Isi_2,Weak_Ising,Su2024,Tirrito2023,Roser2023,Biella2021,Tista2024,Pavigilianti2023}, fermionic systems~\cite{Buchhold2021,Poboiko2024,Muller2022,Chatterjee2024,Minato2022,Jin2024,Kaijian2021,FT}, many-body localized systems~\cite{MIPT_MBL}, higher-dimensional systems~\cite{Turkeshi2020,geo_2,Nahum2021,geo_4,geo_5} and in experiments with trapped ions and superconducting qubits~\cite{exp_1,koh2023,exp_3,exp_4}.{ This volume-to-area law transition basically arises due to the competition between the unitary evolution, which increases the entanglement, and the process of measurement, which decreases the entanglement. At the same time, ref.~\cite{area_law} shows a measurement-induced phase transition in a one-dimensional quadratic fermionic model under local and probabilistically applied measurement in finite system size and shows in the thermodynamic limit this transition vanishes and only the area law phase persists. }{Absence of MIPT has also been shown for random projective measurements~\cite{ref1}, for open 1D fermion system~\cite{ref2}, for Dirac fermions~\cite{ref3} and in dissipative quantum Ising chain~\cite{ref4}.}

In this work, we shall consider non-random global projective measurement performed on a quantum Ising chain evolving under the transverse Ising Hamiltonian. We start from a chosen pure state $|\psi_0\rangle$ and one time step consists of evolution for time $\tau$ followed by a measurement. (Note that the measurement is performed with certainty at every time step, and the only source of stochasticity is the quantum nature of the system.) At each time-step, we compute three quantities, namely, the bipartite entanglement, the multipartite entanglement, and the survival probability of the initial state $|\psi_0\rangle$. From numerical studies on chains of size $L \le 28$, it is found that (i) the bipartite entanglement shows a volume-to-area law transition at some critical value $\tau=\tau_c$ and (ii) the multipartite entanglement as well as the survival probability also bear the signature of a transition at the same value of $\tau$. Thus, we observe a `measurement-induced phase transition' just as in the case of random circuit models \cite{Skinner,koh2023}. The central result of this work is that, exploiting the analytic nature of the Hamiltonian, we can calculate the survival probability at large $L$ ($\sim 1000$) too and show analytically (and numerically) that the critical value $\tau_c$ scales with system size as $1/\sqrt{L}$. This proves that $\tau_c$ vanishes at the thermodynamic limit, and that the transition we observe at small size exists for finite sizes only. Hence, whether the transition in random circuit or other {similar models} survive in the thermodynamic limit is an important question. 

Incidentally, we also find that in contrast to the random circuit models, our measurement does not always reduce the entanglement, and the transition occurs despite this fact. Furthermore, the transition point seems to exist irrespective of the presence of order in the ground state of the Hamiltonian; we present here the results for $h=1/2$ in the main paper (where the ground state is ordered) and include the (qualitatively similar) results for $h=3/2$ (where the ground state is disordered) and {$h=1$,} (equilibrium transition point) in the Supplementary Material~\ref{supx}. Therefore, the equilibrium critical point has no effect on our observation. {We believe that our protocol can be implemented in such experimental setups, where global measurement is performed in multiparty entangled systems. One such experimental work is demonstrated in Ref.~\cite{MIPT_EPX}, where global { measurement } is performed in a highly entangled system.}

We shall first present the measurement protocol and then the numerical results for the survival probability, bipartite entanglement and generalised geometric measure. Then we shall present the analytic results, followed by summary. 

 \begin{figure}
     \centering
     \begin{subfigure}[b]{0.45\textwidth}
         \centering
    \includegraphics[width=\textwidth]{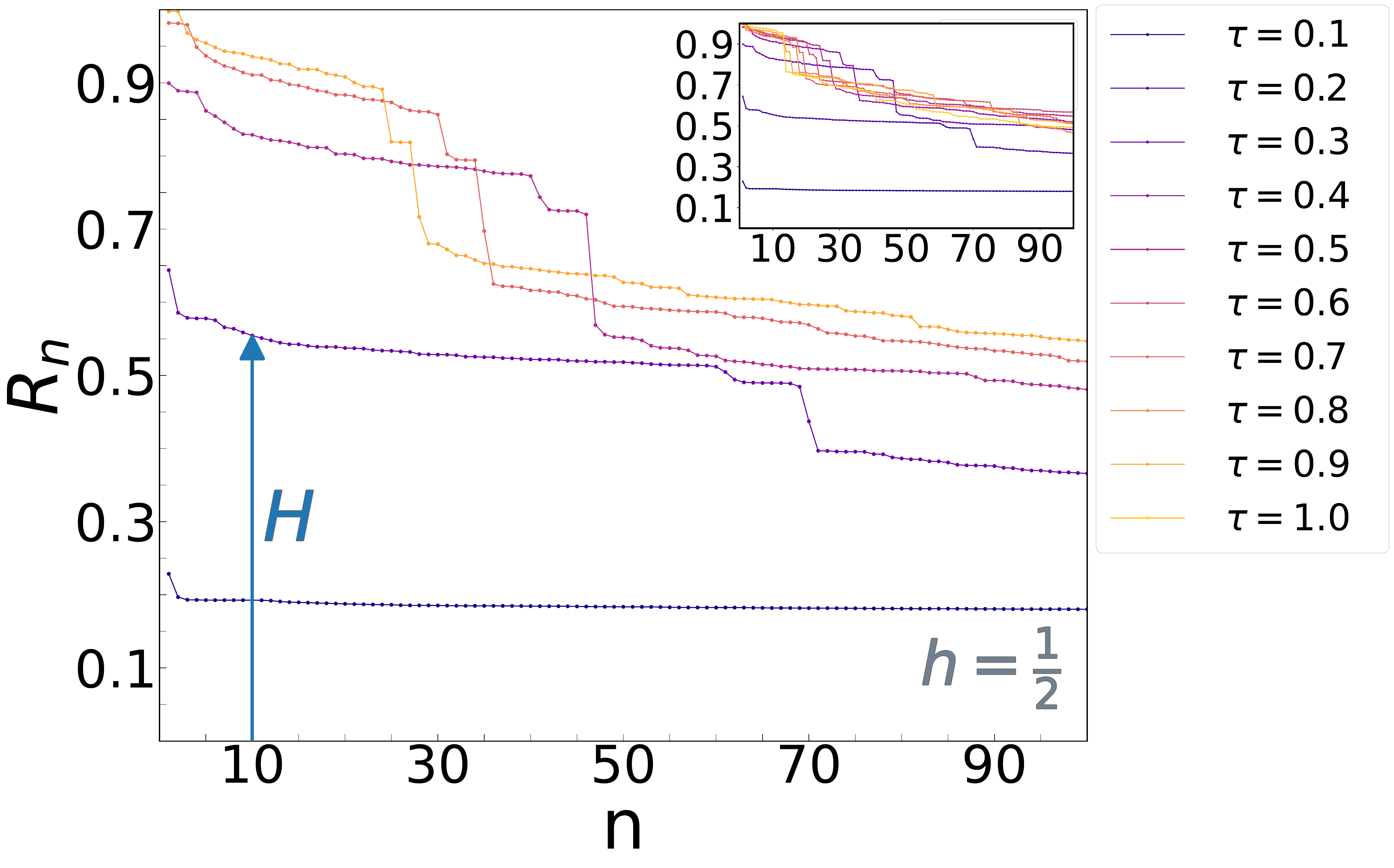}
         \caption{}
         \label{prob_1}
     \end{subfigure}
     \hfill
     \begin{subfigure}[b]{0.35\textwidth}
         \centering
         \includegraphics[width=\textwidth]{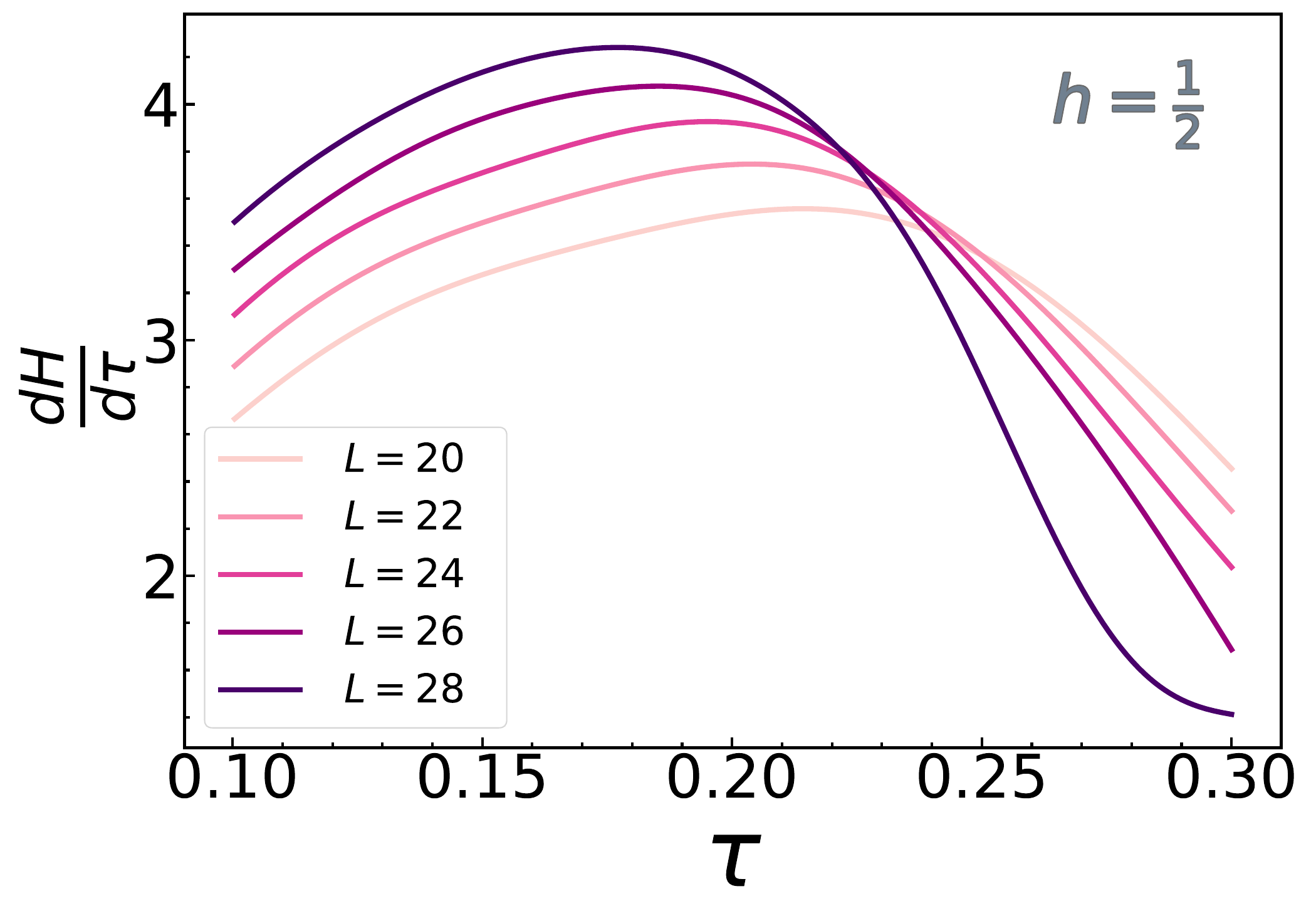}
         \caption{}
         \label{prob_2}
     \end{subfigure}
      \caption{(a) Survival probability $R_n$ (as defined in Eq.~\ref{R_def}) as a function of time-step $n$ for different $\tau$ values at $L=28$ and $h=\frac{1}{2}$. {In the main figure, we consider $\tau$ values ranging from $\tau=0.1$ to $\tau=0.5$, taken at intervals of $0.1$. The inset shows same $R_n$ vs. $n$ plots corresponding to system size $L=28$ for $\tau$ values ranging from $\tau=0.1$ to $\tau=1.0$, in intervals of of $0.1$. The color bar represents the same $\tau$ values in both the main figure and the inset}:(b) shows $dH/d\tau$ vs, $\tau$ plot for time-step $n=10$. The derivative $dH/d\tau$ shows a peak at $\tau=\tau_c$ the location of which varies slightly with system-size and is around 0.2 for {$L=28$}.}
        \label{fig1}
\end{figure}

\noindent
{\bf Protocol and Survival Probability:}\\
We consider (at zero temperature) a chain of $L$ Ising spins in a (pure) state where all the spins are in the +Z direction:

\be |\psi_0\rangle = | 000  \cdots 0 \rangle \ee

\noindent
It is allowed to evolve unitarily for a time $\tau$ under the transverse Ising Hamiltonian
\be   \mathcal{H}= -\sum_{j=1}^{L} s^x_j s^x_{j+1} - h \sum_{j=1}^{L} s^z_j  \label{H_def}  \ee

 We now perform a projective measurement which provides a binary (yes/no) answer to the question ``Are all spins up?''.

 We perform the experiment on a large number ${\mathcal N}$ of identically prepared replicas, then note down the fraction of cases where the answer is {\em yes} and call this fraction $p_1$. We discard the replicas that gave answer {\em yes} and with the other replicas carry on the unitary evolution under ${\mathcal H}$ for another time interval $\tau$. Then we perform the same measurement and note down the number of replicas, say, $p_2{\mathcal N}$, which yield {\em yes}. We repeat this procedure of unitary evolution, followed by projective measurement, so that after the $n$-th cycle, we get the number of replicas, say, $p_n{\mathcal N}$, which yield {\em yes}. 
 Then the probability that the system has yielded {\em no} for the first $n$ successive measurements is
\be R_n \equiv 1- \sum_{k=1}^n p_k \label{R_def}. \ee
 We shall call $R_n$ the survival probability as the initial state has escaped detection over $n$ steps with this probability. The above protocol for dynamics has been used earlier~\cite{SDG_1,SDG_2,SDG_3} as a reasonable protocol for measuring the probability of first occurrence within the framework of quantum measurement.\\
 
 After the time evolution from $t=0$ to $t=\tau$, the system attains a state $\exp(-i\mathcal{H}\tau) |\psi_0\rangle$ and the effect of the measurement is to project out  $|\psi_0\rangle$ and get the state
\be |\psi_1\rangle = e^{-i\mathcal{H}\tau}| \psi_0\rangle - \langle \psi_0 | e^{-i\mathcal{H}\tau} |\psi_0\rangle  |\psi_0\rangle \ee
at the completion of the first time step. Now the state $|\psi_1\rangle$ evolves again for time $\tau$ under $\mathcal{H}$, undergoes measurement, and becomes
\be |\psi_2\rangle =  e^{-i\mathcal{H}\tau} |\psi_1\rangle - \langle \psi_0 | e^{-i\mathcal{H}\tau} |\psi_1\rangle  |\psi_0\rangle \ee
Following this principle, the state function $|\psi_n\rangle$ after $n$ steps can be calculated numerically. We have used the Chebyshev polynomial technique~\cite{Chev.} to facilitate the computation and studied entanglement and survival probability for the state $|\psi_n\rangle$.\\

The plot of $R_n$ as a function of $n$ (for a given $\tau$) shows a plateau region at small $n$. As we increase the value of $\tau$, the height of this plateau approaches unity (Fig.~\ref{prob_1}). The principal quantity of our study is {\em the rate at which the height of the plateau region varies with $\tau$} (Fig.~\ref{prob_1}). 
{ The precise definition of this height is as follows. The $R_n$ vs $n$ curve shows at least one range of values of $n$ (say $n_1<n<n_2$) where the plot is horizontal and $R_n$ does not vary with $n$. The value of $R_n$ in this region is identified as the height of the plateau. In some cases there are several such regions and then we choose the first region (that is, the one for which $n_1$ and $n_2$ are smallest) for measuring the height of the plateau. We describe the height (H) of the plateau pictorially in Fig.~\ref{prob_1}.} We observe that the derivative of this height with respect to $\tau$ shows a peak at some value of $\tau$ (Fig.~\ref{prob_2}). Thus, for equispaced values of $\tau$, the vertical separation of the plateau regions is larger for intermediate values of $\tau$.
We note that the height of the peak rises with system size and identify this value of $\tau$ as the critical value $\tau_c$ (Fig.~\ref{prob_2}). We shall now see that at $\tau_c$ the entanglement entropy also shows a typical transition.

\noindent
{\bf Bipartite Entanglement:} \\
The system of $L$ sites is divided into two subsystems A and B, consisting of $\ell$ and $L-\ell$ (consecutive) sites. The entanglement entropy for a pure state is computed as 
\be S(n,\ell) = -\Tr[\rho_A \ln \rho_A] = -\Tr[\rho_B \ln \rho_B], \ee
where $\rho_A$ and $\rho_B$ are the reduced density matrices for the subsystems $A$ and $B$.  
In the context of MIPT, a volume law to area law transition of the entanglement entropy at some critical rate of measurement ($\mathcal{P}_c$) is identified {\cite{Skinner} by using a scaling relation \[S(\mathcal{P})-S(\mathcal{P}_c)=F\left((\mathcal{P}-\mathcal{P}_c) L^{1/\nu}\right)\]
Choosing $\ell=L/4$, and using a scaling function
\be S(\tau)-S(\tau_c)=F\left((\tau-\tau_c) L^{{1/\nu}}\right) 
\label{collapse1}\ee
{In Fig.~\ref{bipartite}~(a)), considering various $\tau$ values, we observe
that the curves remain nearly flat for $\tau < 0.20$, while for $\tau > 0.20$ a linear growth of entanglement emerges. This behavior
clearly indicates a transition from an area-law to a volume-law scaling, occurring
around ${\tau_c \approx 0.20}$.}
we also find a collapse (Fig.~\ref{bipartite}~(b)) with $\tau_c=0.2$ and {$\nu=1.013$}, thus, confirm the presence of an area-to-volume law transition more robustly. A similar study based on bipartite entanglement entropy for $h=\frac{3}{2}$ and $h={1}$  is provided in the supplementary material~\ref{supx}, an area-to-volume law transition occurs about $\tau_c=0.1$ and $\tau_c=0.16$ for $h=\frac{3}{2}$ and $h={1}$ respectively. An additional confirmation for the transition is observed from the collapse of the plot of $H(\tau) - H(\tau_c)$ vs $(\tau - \tau_c)L^{ 1/\lambda}$ (Fig. \ref{h_scaling}), where $\lambda=0.4317$. 

\begin{figure}
    \centering
    \includegraphics[width=1\linewidth]{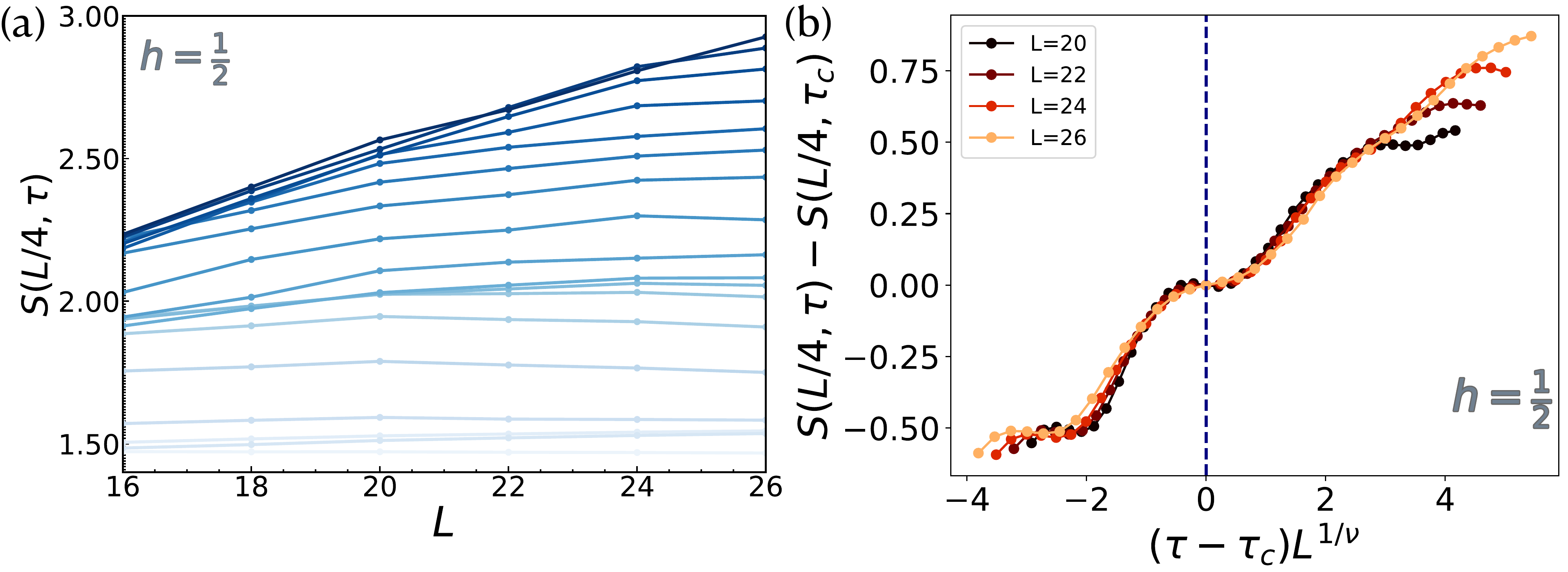}
    \caption{Behaviour of the bipartite entanglement entropy with system size $L$ and $\tau$ for $h=1/2$. In panel (a), entanglement entropy for the whole system in the $L/4$ : rest bipartition ($S(L/4)$) is plotted for different $\tau$ values with respect to the system size $L$. The $\tau$ values increase as the color of the curves becomes deeper. In this context, we have considered values of $\tau$ ranging from $0.06$ to $0.40$, with an increment of $0.02$ between successive $\tau$ values. We have taken $n=10$ for measuring $S$. {Now, among all the $\tau$ values, i.e for $\tau=0.06,0.08,0.10,0.12,0.14,....,0,20,0.22,0.24...,0.40$, the curves are almost flat bellow $\tau=0.20$ and above $\tau=0.20$, almost linear increment of entanglement starts to arise. It is apparent that there is an area law to volume law transition {around $\tau_c=0.20$}.} In panel (b) we have shown the scaling collapse (following Eq. \ref{collapse1}) of the entanglement curves occurs when $\tau_c=0.2$ with exponent {$\nu=1.013$}.}
    \label{bipartite}
\end{figure} 

\begin{figure}
    \centering
    \includegraphics[width=0.70\linewidth]{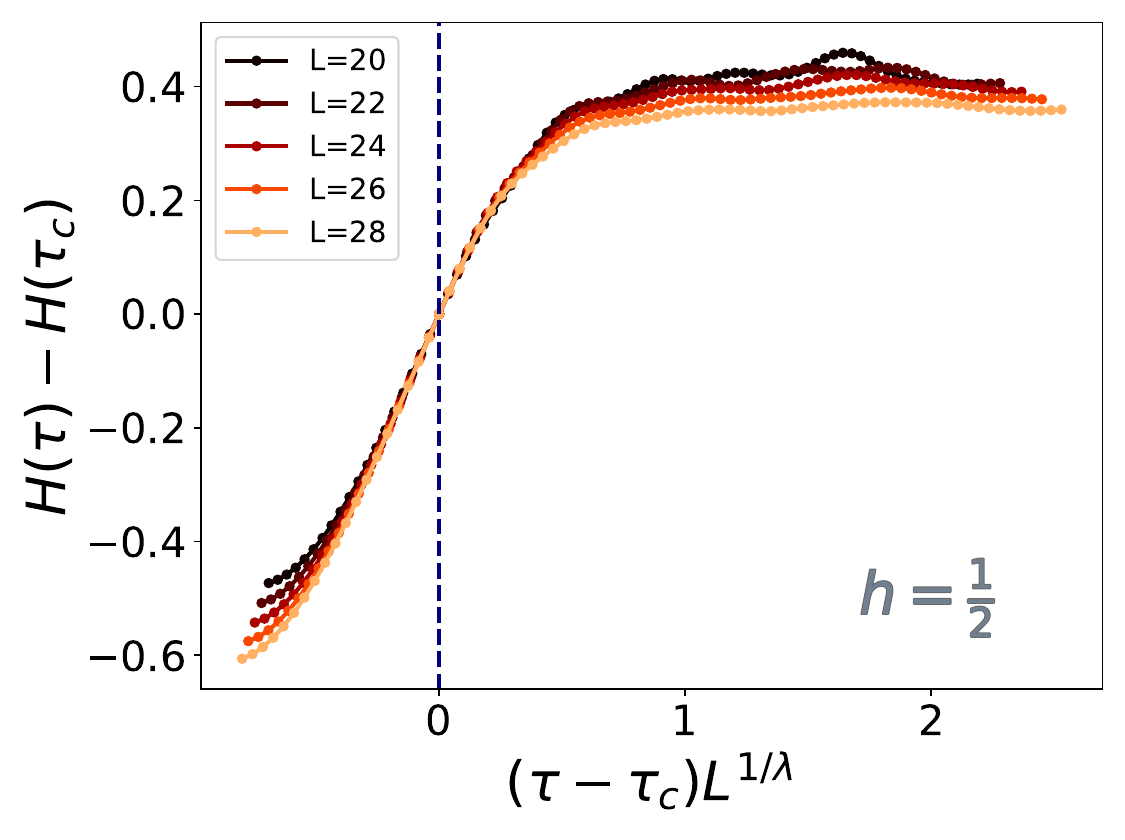}
    \caption{Scaling collapse of height of plateau region in $R_n$ vs $n$ curve at $\tau_c=0.2$ for different system sizes at the field strength $h=1/2$. The value of {$\lambda$} corresponding to height collapse is given by, {$0.4317$}.}
    \label{h_scaling}
\end{figure} 
\begin{figure*}
    \centering
    \includegraphics[
        width=17.4cm,height=4.8cm,
    ]{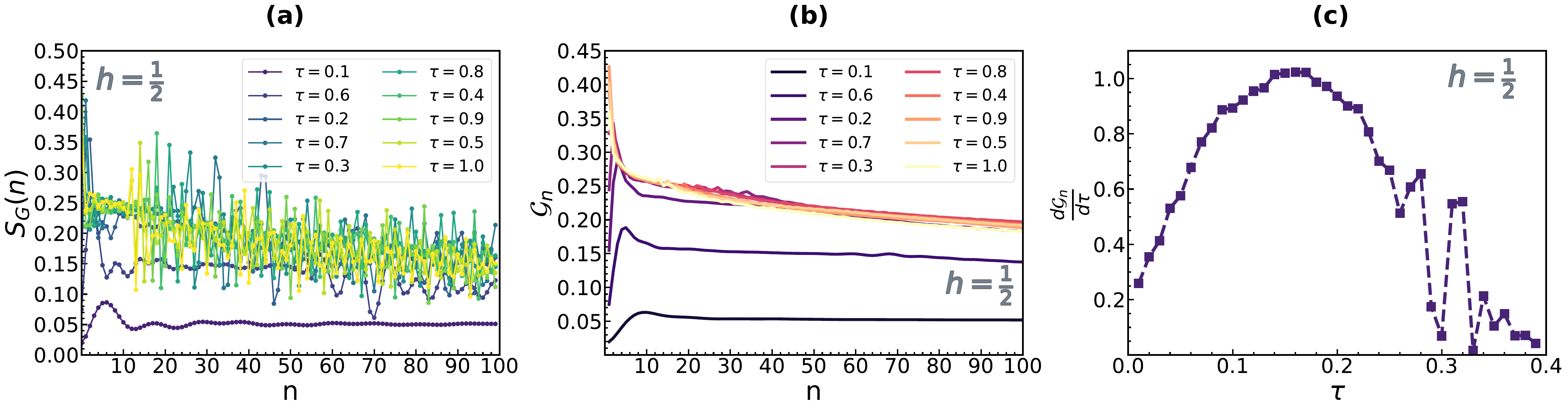}
    \caption{Variation  of multipartite entanglement with time-step $n$ and $\tau$ for $h=\frac{1}{2}$: In panel (a), we have plotted $S_G(n)$ vs $n$ for different $\tau$ values. The curves for $\tau>0.2$ are nearly coincident, but are separated from the one for $\tau=0.2$. In panel (b), $\mathcal{G}_n$ vs $n$ is plotted at $L=26$, where $\mathcal{G}_n$ is the cumulative average, namely the average of the sequence $\bigl\{S_G(1),S_G(2),S_G(3),S_G(4)...S_G(n)\bigl\}$. In panel (c) we have plotted $d\mathcal{G}_n/d\tau$ vs $\tau$, where the peak occurs around $\tau=0.2$, indicating that $\tau_c=0.2$. }
   
    \label{prob}
\end{figure*}

\noindent
{\bf Generalized Geometric Measure:}\\
For quantifying multipartite entanglement, we utilize the measure referred to as the generalized geometric entanglement (GGM)~\cite{GGM_1,GGM_2,GM_2,GM_3,GM_31,GM_4,GM_5,GM_6}. The GGM of a pure $L$-partite state, \(|\psi_n\rangle\) is defined as
\begin{equation}\label{eq_9}
G_n = 1 - \Lambda^2_{max}(\ket{\psi_n}),
\end{equation}
where $\Lambda_{max}(\ket{\psi_n})$ is the maximum fidelity between the given state $\ket{\psi_n}$ and an arbitrary non-genuinely multiparty entangled state. This can be written in a closed form~\cite{GGM_1} as
\be G_n = 1 - \max_{\ell} \left\{ \lambda_{\ell, L-\ell}^2 \right\}_l \ee
where $\lambda_{\ell, L-\ell}$ is the maximal Schmidt coefficient in the bipartite split $\ell:L-\ell$ of our chain.
Since we calculate the GGM for the state $|\psi_n\rangle$ after the $n$th measurement, we plot the {\em stochastically attained GGM}~(SAG), defined as}
\[ S_G (n)  = R_n G_n \]
The plot of $S_G(n)$ with $n$ at field strength $h=1/2$ as depicted in the panel (a) of Fig.~\ref{prob} for equispaced $\tau$ values shows the presence of plateaus (just as in the case of survival probability) in spite of strong fluctuations. Also, the plateaus are found to be more separated (vertically) near $\tau_c$. To avoid fluctuations, we perform cumulative averaging of the data for each $\tau$ value (panel (b) of Fig.~\ref{prob}). The cumulative average for a sequence, $\bigl\{x_1, x_2, x_3, \ldots, x_n\bigl\}$, is defined as $\mathcal{M}_i=\frac{1}{i} \sum_{k=1}^{i} x_k$ for $i=1, 2, \ldots n$. Using the cumulative average, we plot (panel (c) of Fig.~\ref{prob}) the derivative of the height of the plateau region and find a peak at $\tau_c=0.2$, which we identify as the transition point.

Thus, for sizes $L\le 28$, the three quantities, namely survival probability, bipartite entanglement entropy, and genuine multipartite entanglement, show a measurement-induced transition at a common value of $\tau_c$ around $0.2$ at field strength $h=1/2$.

\noindent
{\bf Recursion Relation for Survival Probability:}\\
 Although we could not derive a closed-form expression for the survival probability, it is possible to derive a recursion relation that enables one to compute the probability $R_n$ for large system-size. To this end,
 we introduce the quantities
\be |\phi_n\rangle = e^{-i\mathcal{H}\tau n} |I\rangle, \;  f_n = \langle I |e^{-i\mathcal{H}\tau n} |I\rangle,\;  n=0,1,2, \cdots \ee
Then, $|\phi_0\rangle =  |I\rangle$, $f_n=\langle I|\phi_n\rangle$, $\langle \phi_m|\phi_n\rangle = f_{n-m}$ and 
\be |\psi_1\rangle = |\phi_1\rangle - f_1|I\rangle \ee
After the second time-step, 
\bea |\psi_2\rangle & = &  e^{-i\mathcal{H}\tau} |\psi_1\rangle - \langle I | e^{-i\mathcal{H}\tau} |\psi_1\rangle  |I\rangle \nonumber \\
&=& |\phi_2\rangle  - f_1|\phi_1\rangle + (f_1^2 - f_2)|I\rangle \eea
The basic idea is to observe that for $n=0, 1, 2$, the wave-function can be expressed as
\be |\psi_n\rangle = \sum_{m=0}^n C_m^{(n)} |\phi_m\rangle \ee
and that the next wave-function 
 \be |\psi_{n+1}\rangle = e^{-i\mathcal{H}\tau} |\psi_n\rangle - \langle I | e^{-i\mathcal{H}\tau} |\psi_n\rangle  |I\rangle \ee
 can also be written as,
\be |\psi_{n+1}\rangle = \sum_{m=0}^{n+1} C_m^{(n+1)} |\phi_m\rangle \ee 
with the recursion relations
\be C^{(n+1)}_0=-\sum_{m=0}^n C^{(n)}_mf_{m+1}, \;\;\;  C^{(n+1)}_m=C^{(n)}_{m-1}  \label{recursion1} \ee
for $0<m\le n$.
Note that
\bea C_n^{(n)} & = & C^{(n+1)}_{n+1} = 1, \label{recursion2} \\ 
C_0^{(1)} & = & -f_1, \;\;\;  C_1^{(2)} = -f_1, \;\;\;  C_0^{(2)} = f_1^2 - f_2 \label{recursion3} \eea
and that
\be C_m^{(n)} = C_{m-1}^{(n-1)} = C_0^{(n-m)} \label{recursion4} \ee
(Similar recursion relations has been discussed extensively \cite{Barkai2017,Barkai2024} in a somewhat different context.)
Using the recursion relations, one can find the survival probability after $n$ measurements as
\be R_n = \langle \psi_n | \psi_n \rangle = \sum_{m_1, m_2 =0}^{(n)} \left(C_{m_1}^{(n)}\right)^* C_{m_2}^{(n)} \, f_{m_2-m_1} \label{RnC} \ee

This expression is valid for any Hamiltonian, provided $f_n$ are known. For the specific case of transverse Ising Hamiltonian, one can derive an expression for $f_n$ in closed form for any even value of $L$, by using the exact solution~\cite{Damski} (See Supplementary Material)
\be f_n = \prod_k \left[\cos (\lambda_k n \tau) + i\sin(\lambda_k n \tau) \cos(2\theta_k) \right]  \label{fn_final} \ee
where $\lambda_k$ and $\theta_k$ are defined by
\[ \lambda_k = 2\sqrt{h^2 +1 +2 h \cos k}, \;\;\; e^{2i\theta_k} = \frac{2(h+e^{-ik})}{\lambda_k}\]
In principle, one  should be able to calculate the survival probability for any $L$ and $n$ using Eqs. (\ref{recursion1}, \ref{RnC}, \ref{fn_final}), but due to some precision problems we could not go beyond $L=1000$ and $n=10000$. \\
As for small size, the survival probability shows a plateau region and the derivative of the height $H$ of the plateau with respect to $\tau$ shows a peak at a critical point $\tau=\tau_c$. (Fig.~\ref{prob_l}). It is found that if one uses instead of $\tau$ a scaled variable, $\sigma=\tau\sqrt{L}$ then the plot of $dH/d\sigma$ vs $\sigma$ coincide for different values of $L$. This indicates that the position of the peaks $\tau_c$ varies as $1/\sqrt{L}$ and vanishes for large $L$.  One can derive analytically this scaling law in the limit of small $\tau$ and small $n\tau$ (see Supplementary ~\ref{supx}). Thus, the transition exists only for finite $L$ and vanishes in the thermodynamic limit. We would like to add that scaling of $\tau_c$ on the basis of small-size data does  not give this $1/\sqrt{L}$ scaling; rather, $|\tau_c(L) - \tau_c(\infty)|$ scales as $\log L$. The detailed description is provided in supplementary materials~\ref{supx}.

\begin{figure}
     \centering
     \begin{subfigure}[b]{0.24\textwidth}
         \centering
         \includegraphics[width=\textwidth]{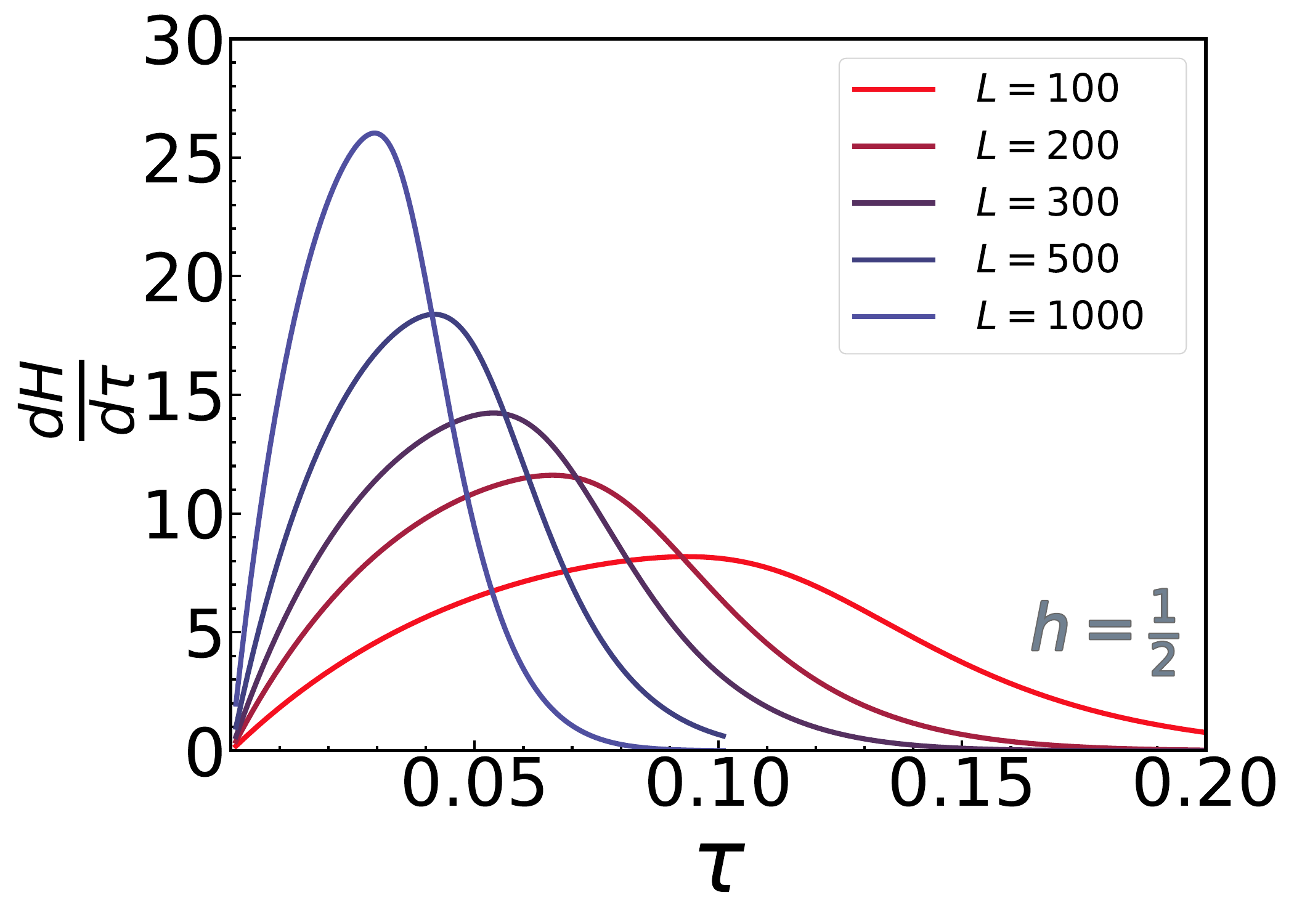}
         \caption{}
         \label{prob_l_1}
     \end{subfigure}
     \begin{subfigure}[b]{0.23\textwidth}
         \centering
         \includegraphics[width=\textwidth]{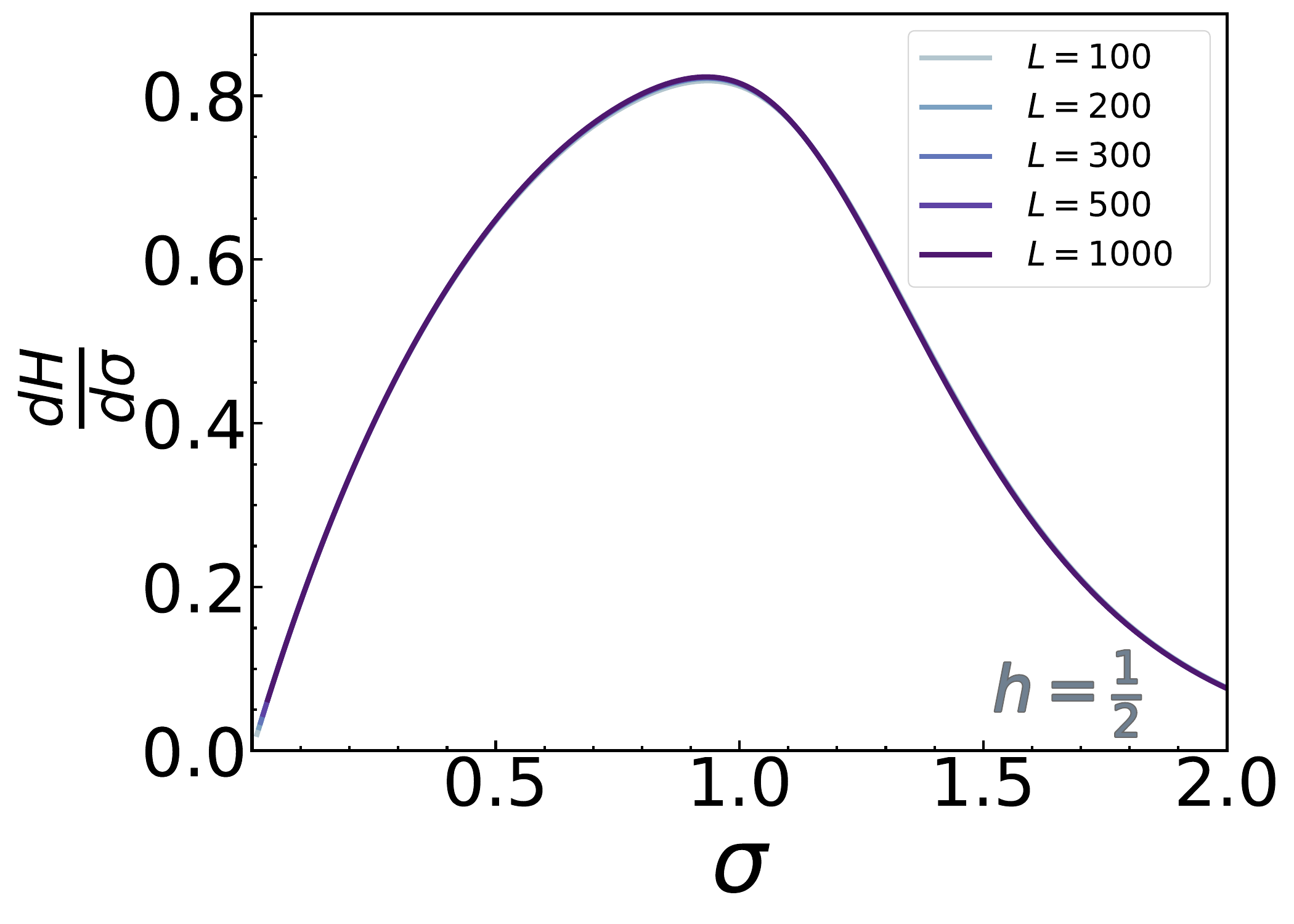}
         \caption{}
         \label{prob_l_2}
     \end{subfigure}
      \caption{Survival probability at large system sizes. (a) :  Plot of $dH/d\tau$ vs $\tau$, where it is clear that the peak at $\tau_c$ moves towards $0$ as $L$ increases. (b): Plot of $dH/d\sigma$ vs. $\sigma$ for various system sizes at field $h=\frac{1}{2}$ where $\sigma=\tau \sqrt{L}$. All the curves coincide and show a peak at $\sigma=1$, indicating the scaling $\tau_c \sim 1/\sqrt{L}$.}
        \label{prob_l}
\end{figure}

There are some comments (1) For quantum circuits, the transition arises  from a competition  between the measurement process, which decreases the entanglement and the unitary evolution, which increases the entanglement. In contrast, for our case the transition arises in spite of the fact that in some cases the measurement increases the entanglement, while in some cases it decreases (Fig.~S.9 of the supplementary materials~\ref{supx}). (2) The end of the plateau region in the $R_n$ vs $n$ plot is accompanied by a sudden dip in the entanglement (Fig.~S.10 of the supplementary materials~\ref{supx}). Thus, the presence of the plateau in the $R_n$ vs $n$ plot seems to be related to the entanglement entropy. 
(3) As mentioned earlier, the critical value $\tau_c$ decreases with system size. Numerical results show that it also depends on the transverse field $h$ (see Supplementary Materials for details). (4)  Normally, the MIPT observed in quantum circuits is independent of the parameters of the Hamiltonian. But in our case, we observe that the survival probability at long time shows a logarithmic decay for $h=3/2$, when the ground state of Hamiltonian is disordered (see Supplementary Materials).
(5) One must note that all the results discussed above are for the initial state where all the spins are pointed up in the Z direction. {We also see for the initial Neel state and field strength $h=\frac{1}{2}$, there exists phase transition, although for initially oriented spins along the $X$ direction, the area-to-volume law transition is observed by bipartite entanglement but is not captured by survival probability.} 

\noindent 
{\bf Discussion:}\\
We present the observation of measurement-induced phase transition in a quantum spin model with projective non-random measurement. We start with a chain of $L$ Ising spins all oriented in the +Z direction. In each time step, it evolves unitarily under the transverse Ising Hamiltonian for time $\tau$, and then a measurement is performed that gives a binary answer to the question, ``Is the magnetic moment per site equal to 1?". After $n$ such time-steps, the probability that the answer is ``no'' in each case is termed as the survival probability $R_n$. We have derived a recursion relation, using which we could compute $R_n$ for size up to $L=1000$. The plot of $R_n$ vs. $n$ shows a plateau region at small $n$. The key parameter is the rate at which the height (say $H$) of this plateau changes with $\tau$. The plot of $\frac{dH}{d\tau}$ shows a peak at some value of $\tau$, which we identify as the critical point $\tau_c$, since the bipartite and multipartite entanglement undergoes an area-law to volume-law transition there for small size. We prove analytically that, for small $\tau$, the critical point $\tau_c$ scales as $1/\sqrt{L}$. On using the scaled variable $\sigma=\tau \sqrt{L}$, one finds that the plots of $\frac{dH}{d\sigma}$ vs $\sigma$ for different system sizes are coincident, and as $L$ increases, the peak in the plot of $\frac{dH}{d\tau}$ rises in height and shifts towards zero. Hence, the transition is only observable for finite $L$. { As mentioned in the Introduction, MIPT has been observed numerically in a variety of systems and in different contexts. It will be interesting to explore size-dependence of MIPT in these situations.}\\

\noindent
{\bf Acknowledgements:}\\
PN acknowledges UGC, India for financial support (Ref. No. 191620072523) and Harish Chandra Research Institute for access to their infrastructure. SDG acknowledges Eli Barkai for insightful discussions in the conference titled Quantum Trajectories held in ICTS, India.

\onecolumngrid

\section{Supplemental Material}\label{supx}
\renewcommand{\thefigure}{S.\arabic{figure}}
\setcounter{figure}{0}
{\bf Derivation for $f_n$ in Eq. (21):}\\
As mentioned in the text, for transverse Ising Hamiltonian, one can derive an expression for $f_n$ in closed form for any even value of $L$, by using the known analytic solution~\cite{Damski}.
One can transform spin variables $s_j$ to fermion variables $a_j$ and perform Fourier transformation to fermion variables $a_k$, to get $\mathcal{H}$ as a Kronecker sum of commuting Hamiltonians $\mathcal{H}_k$ :
\bea \mathcal{H}&=&  \sum_{k=0}^{\pi}  \mathcal{H}_k, \;\;
\mathcal{H}_k = - 2 i \sin k \left[ a_k^{\dagger}a_{-k}^{\dagger} + a_k a_{-k} \right] - 2(h + \cos k) \left[ a_k^{\dagger}a_k + a_{-k}^{\dagger}a_{-k} - 1 \right] \eea
where   $k=(2n+1)\pi/L$ with $n=0,1,2,\cdots,L/2-1$. Each of these $\mathcal{H}_k$'s can be described by four basis states namely, $|00 \rangle_k$, $|11 \rangle_k$, $|10\rangle_k$ and $|01\rangle_k$, where the numbers in each basis signify the occupation status of the fermions having momenta $+k$ and $-k$ respectively. (We consider the even-occupation states only.) The ground state and excited state of $\mathcal{H}_k$ are 
\bea |GS \rangle_k & = &  i \cos \theta_k |11\rangle_k - \sin \theta_k |00\rangle_k, \nonumber  \\
|ES \rangle_k  &=&  i \sin \theta_k |11\rangle_k + \cos \theta_k |00\rangle_k \label{eigenstates} \eea
with eigenvalues $\mp\lambda_k = \mp 2\sqrt{h^2 +1 +2 h \cos k}$ respectively. Here, $\theta_k$ is defined by $ e^{2i\theta_k} = 2(h+e^{-ik})/\lambda_k$. To proceed further, we note that since $\sum_j s^z_j = \sum_k(2a_k^{\dagger}a_k -1)$, the state $|0\; 0\; 0 \cdots \rangle$ corresponds to the one with all $k$-modes (positive and negative) occupied. Using Eq.~(\ref{eigenstates}), one then obtains
\begin{eqnarray}
e^{-i\mathcal{H}_k n \tau} |11\rangle_k & = & \left[\cos (\lambda_k n \tau) + i\sin(\lambda_k n \tau) \cos(2\theta_k) \right] |11\rangle_k - \sin(2\theta_k) \sin(\lambda_k \tau) |00\rangle_k 
 \label{1100c} \end{eqnarray}

This gives finally,
\be f_n = \prod_k \left[\cos (\lambda_k n \tau) + i\sin(\lambda_k n \tau) \cos(2\theta_k) \right]  \label{fn_final} \ee

{\bf Derivation of the scaling law $\tau_c \sim 1/\sqrt{L}$}:\\
We first write $f_n$ in Eq.~(\ref{fn_final}) as 
\[ f_n = \rho_n e^{i\Phi_n} \]
and note that
\be \rho_n = \exp \left[ \frac{1}{2}\sum_k \log \left[1 - \left(\frac{2 \sin k \,  \sin (\lambda_k n  \tau)}{\lambda_k}\right)^2 \right] \right] \ee
\be \Phi_n = \sum_k \tan^{-1} \left[ \frac{2(h+\cos k)}{\lambda_k}\, \tan(\lambda_k n\tau)    \right] \ee
For small values of $\tau$ and $n$, one obtains
\[ \Phi_n = \sum_k 2n\tau(h+\cos k) = \mu n\]
where $\mu=2\tau h L$. The recursion relations then tell us,
\[ {\rm Arg}\left(C^{(n)}_m \right) = \mu(n-m) \]
and hence the argument of each summand in the expression for $R_n$ in Eq.~(20) (of main text) is zero. Thus, the main contribution to $R_n$ comes from $\rho_n$, which, in the limit of small  $n\tau$ becomes
\[ \rho_n = \exp\left[ - \frac{1}{2}n^2\tau^2 L\right]  \]
This indicates that the parameters $\tau$ and $L$ may be expected to occur as the combination $\tau^2 L$  in the value of survival probability, leading to the scaling law $\tau \sim 1/\sqrt{L}$. \\

{\bf Dependence of the critical point on system-size and external field:}\\
As discussed in the main text, the location of the critical point $\tau_c$ depends on the size of the system $L$ and the transverse field $h$ as shown in FIG.~\ref{Rn3d}

\begin{figure}[h!]
		\centering
		\includegraphics[scale=0.19]{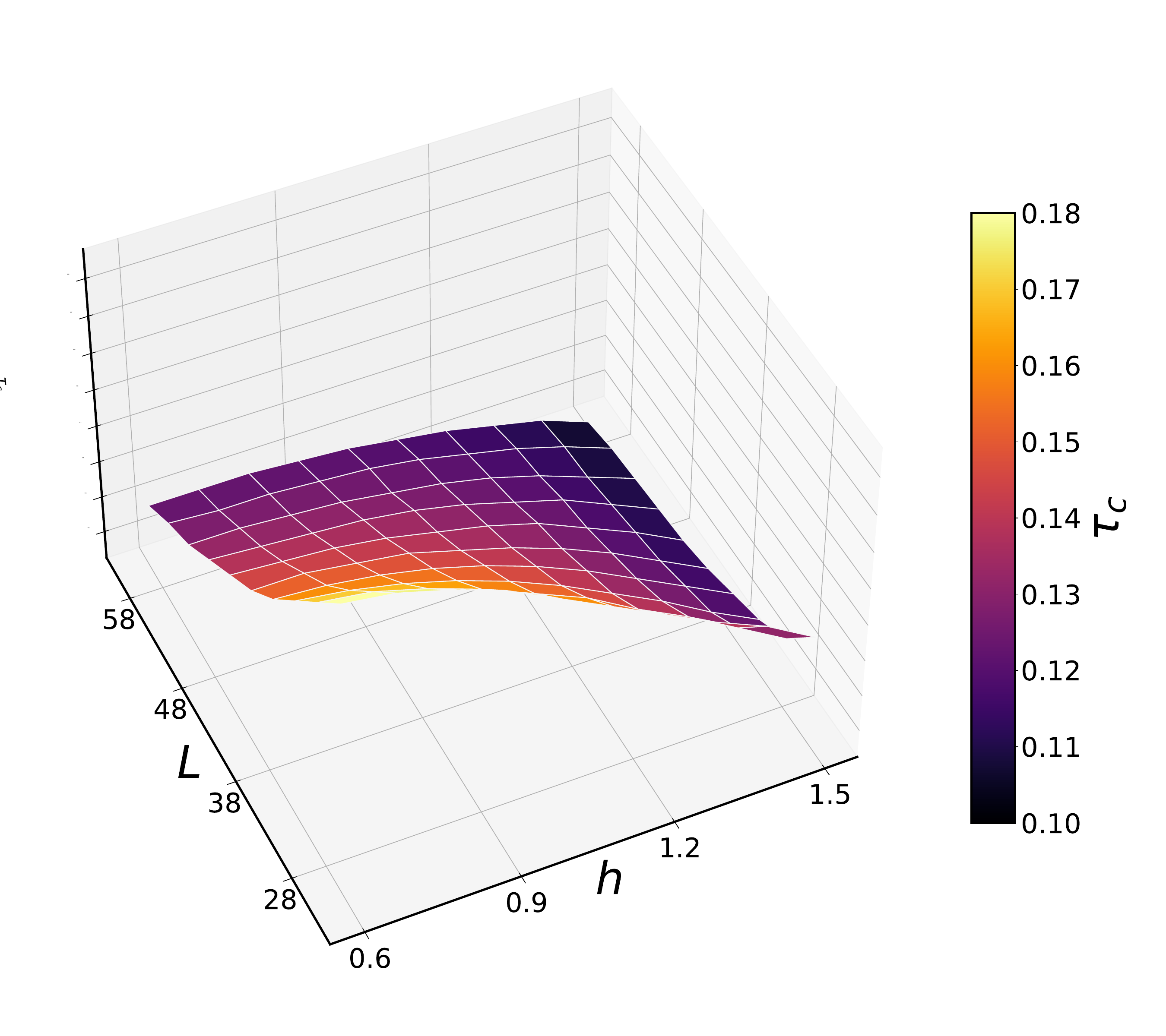}
		\caption{Critical point $\tau_c$ as obtained from the survival probability (see text) as a function of the system size $L$ and field strength $h$.}
  \label{Rn3d}
	\end{figure}

{\bf Dependence of Survival probability decay on transverse field strength:}\\
In FIG (\ref{prob_lo_new}), the survival probability is studied at large $n$ at $h=\frac{3}{2}$ for different $\tau$ values. We can see that $R_n$ decays in logarithmic fashion at large $n$. This kind of logarithmic decay is not present for $h=1/2$. The ground state of transverse Ising Hamiltonian is known to be ferromagnetic for $h<1$  and paramagnetic for $h>1$. Thus, although the MIPT in quantum circuits is independent of the parameters of the Hamiltonian, in our case the observable (namely, the survival probability of the initial state) can detect the ground state of the Hamiltonian.  However, we must mention that when $h$ is too high, such decay is not present even for $h>1$.

\begin{figure*}
     \centering
     \begin{subfigure}[b]{0.40\textwidth}
         \centering
         \includegraphics[width=\textwidth]{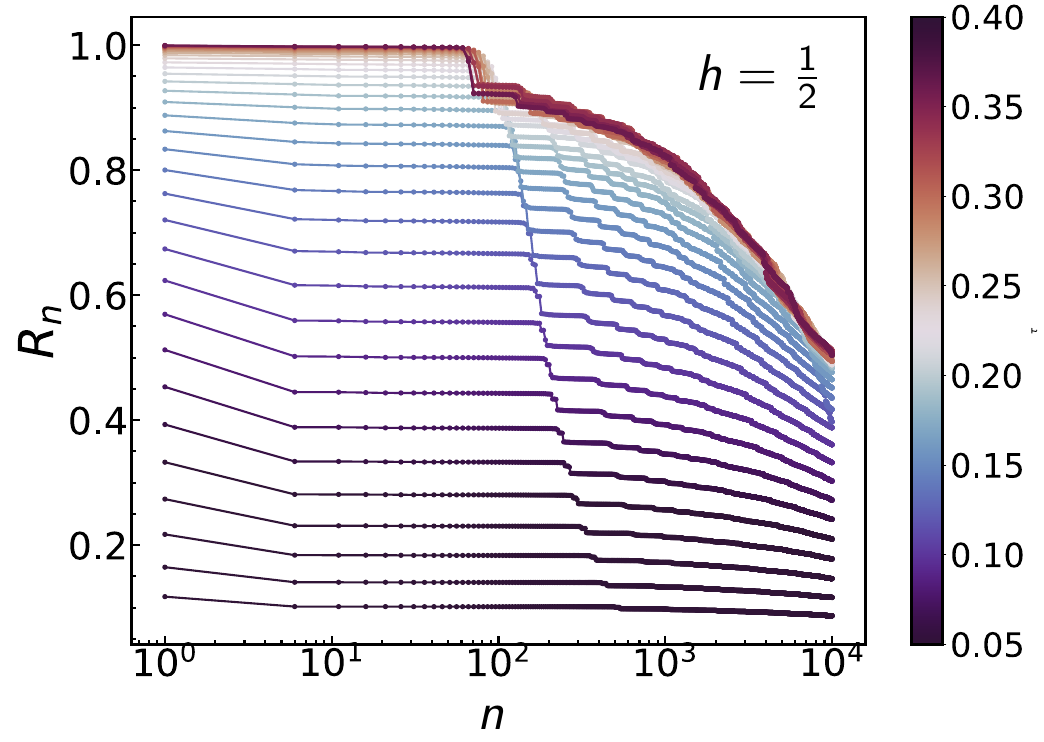}
         \caption{}
         \label{prob_lo_1}
     \end{subfigure}
     \begin{subfigure}[b]{0.40\textwidth}
         \centering
         \includegraphics[width=\textwidth]{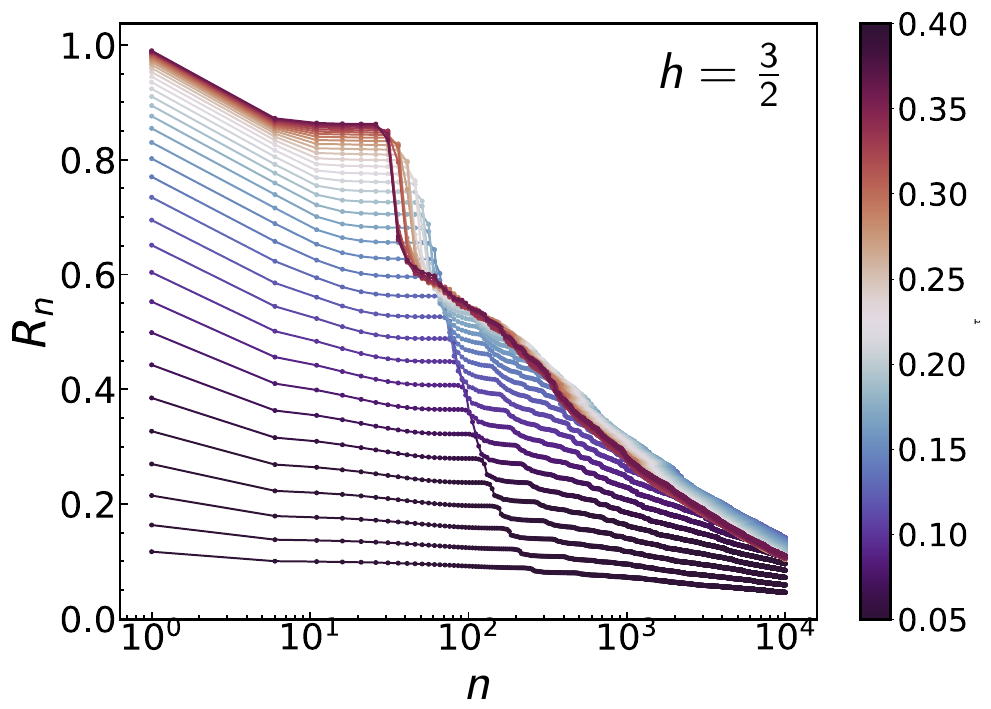}
         \caption{}
         \label{prob_l0_2}
     \end{subfigure}
      \caption{ Survival probability $R_n$ vs time-step $n$ for (a) $h=1/2$ and (b) $h=3/2$  at $L=50$. Note the log-scale along X axis and the linear nature of the curve in panel (b).}
        \label{prob_lo_new}
\end{figure*}

{\bf Results for $h=3/2$:}\\
Survival probability, bipartite entanglement entropy, and multipartite entanglement (stochastically attained GGM) are plotted for $h=3/2$ in Figs. \ref{figp4n}, \ref{bipartitec1}, \ref{SM_GGM} respectively. For large size, the plot of $dH/d\tau$ vs $\tau$ and $dH/d\sigma$ vs $\sigma$ is presented in Fig. \ref{prob_lof}

\begin{figure*}
     \centering
     \begin{subfigure}[b]{0.46
     \textwidth}
         \centering
         \includegraphics[width=\textwidth]{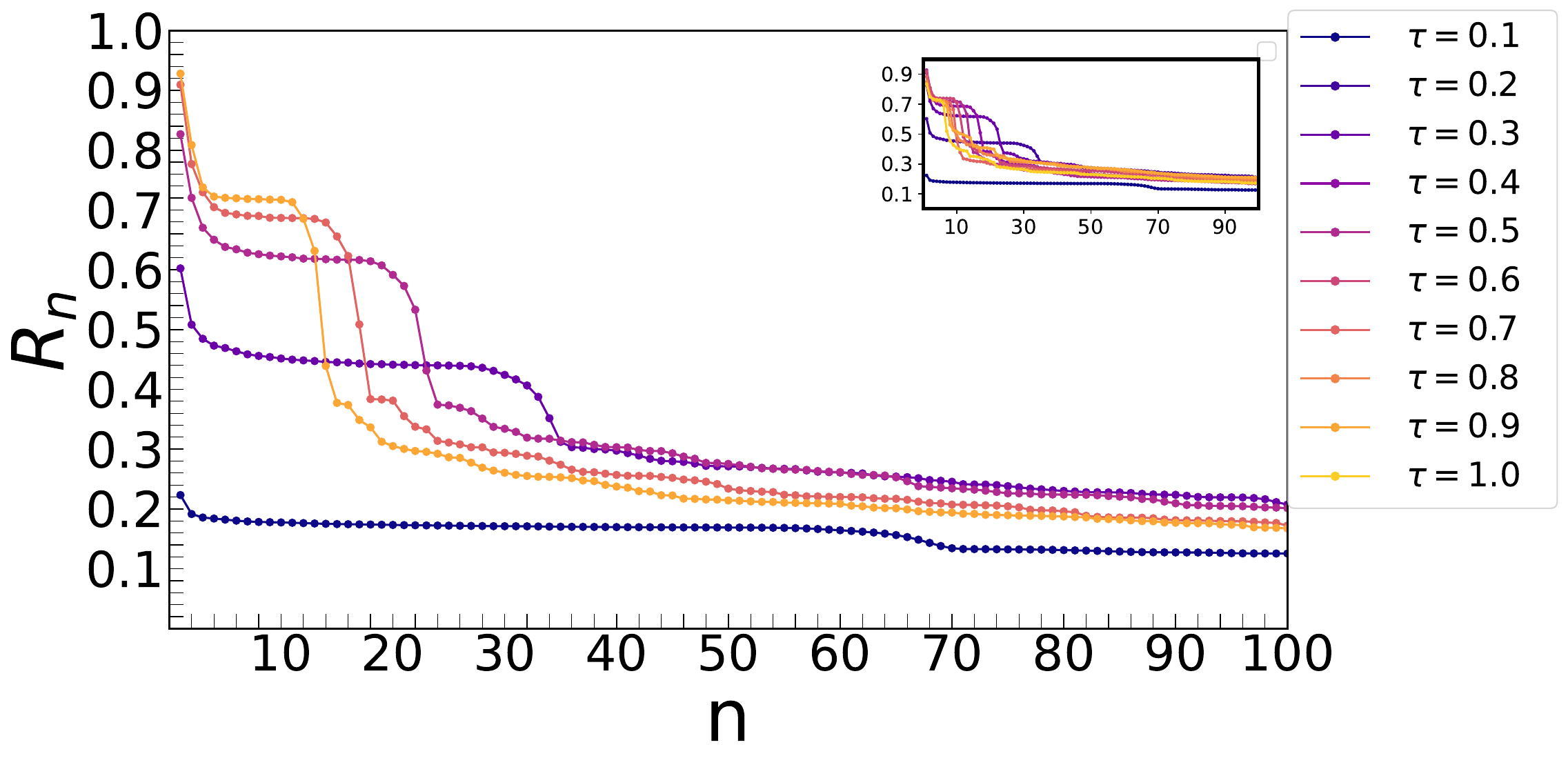}
         \caption{}
         \label{prob_lo_1}
     \end{subfigure}
     \begin{subfigure}[b]{0.32\textwidth}
         \centering
         \includegraphics[width=\textwidth]{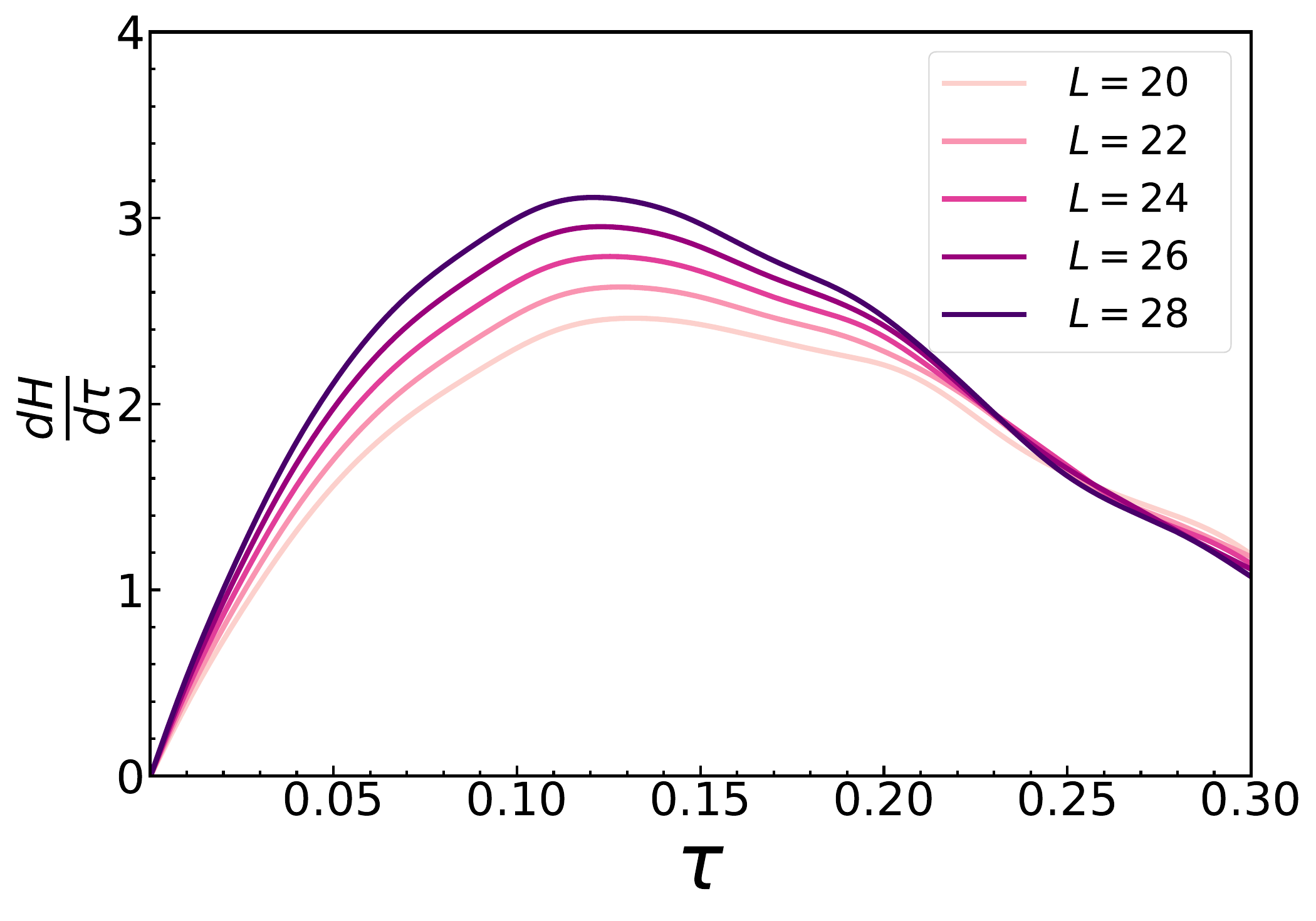}
         \caption{}
         \label{prob_l0_2}
     \end{subfigure}
      \caption{  Survival probability vs time-step and $\frac{dH}{d\tau}$ vs $\tau$ plot. in panel (a) Survival probability vs time-step for different values of $\tau$ at $h=3/2$, $L=28$. {In the main figure, we consider $\tau$ values ranging from $\tau=0.1$ to $\tau=0.5$, taken at intervals of $0.1$. At the same time the inset shows same $R_n$ vs. $n$ plots corresponding to system size $L=28$ for $\tau$ values ranging from $\tau=0.1$ to $\tau=1.0$, in intervals of of $0.1$. The color bar represents the same $\tau$ values in both the main figure and the inset.} In, (b) $\frac{dH}{d\tau}$ vs $\tau$ plot for different system sizes at field $h=\frac{3}{2}$. The transition point is identified as the peak in each curve.}
        \label{figp4n}
\end{figure*}

  \begin{figure}
     \centering
     \begin{subfigure}[b]{0.29\textwidth}
         \centering
\includegraphics[width=\textwidth]{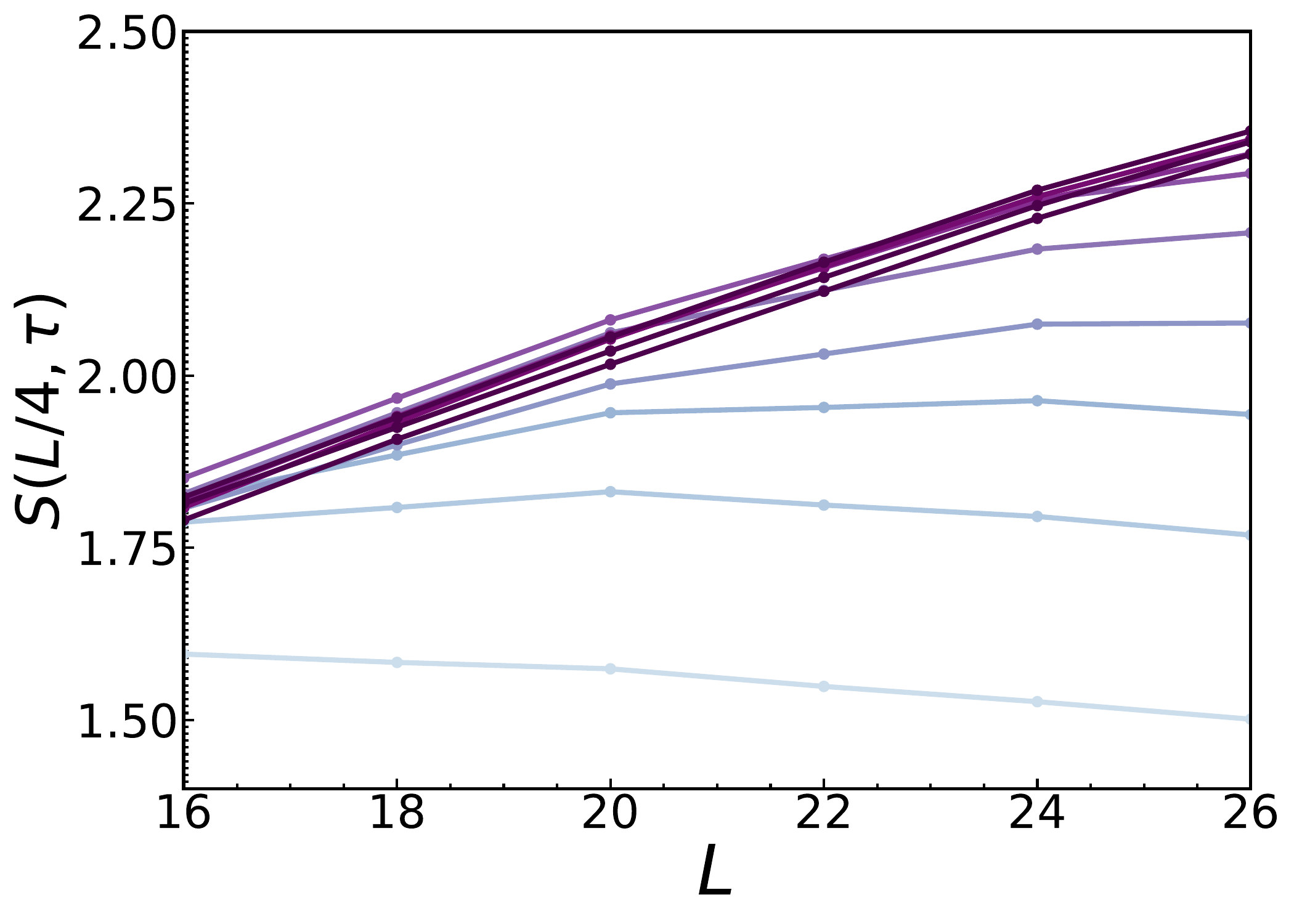}

         \caption{}
         \label{Bi_1}
     \end{subfigure}
        \hfill
     \begin{subfigure}[b]{0.29\textwidth}
         \centering
         \includegraphics[width=\textwidth]{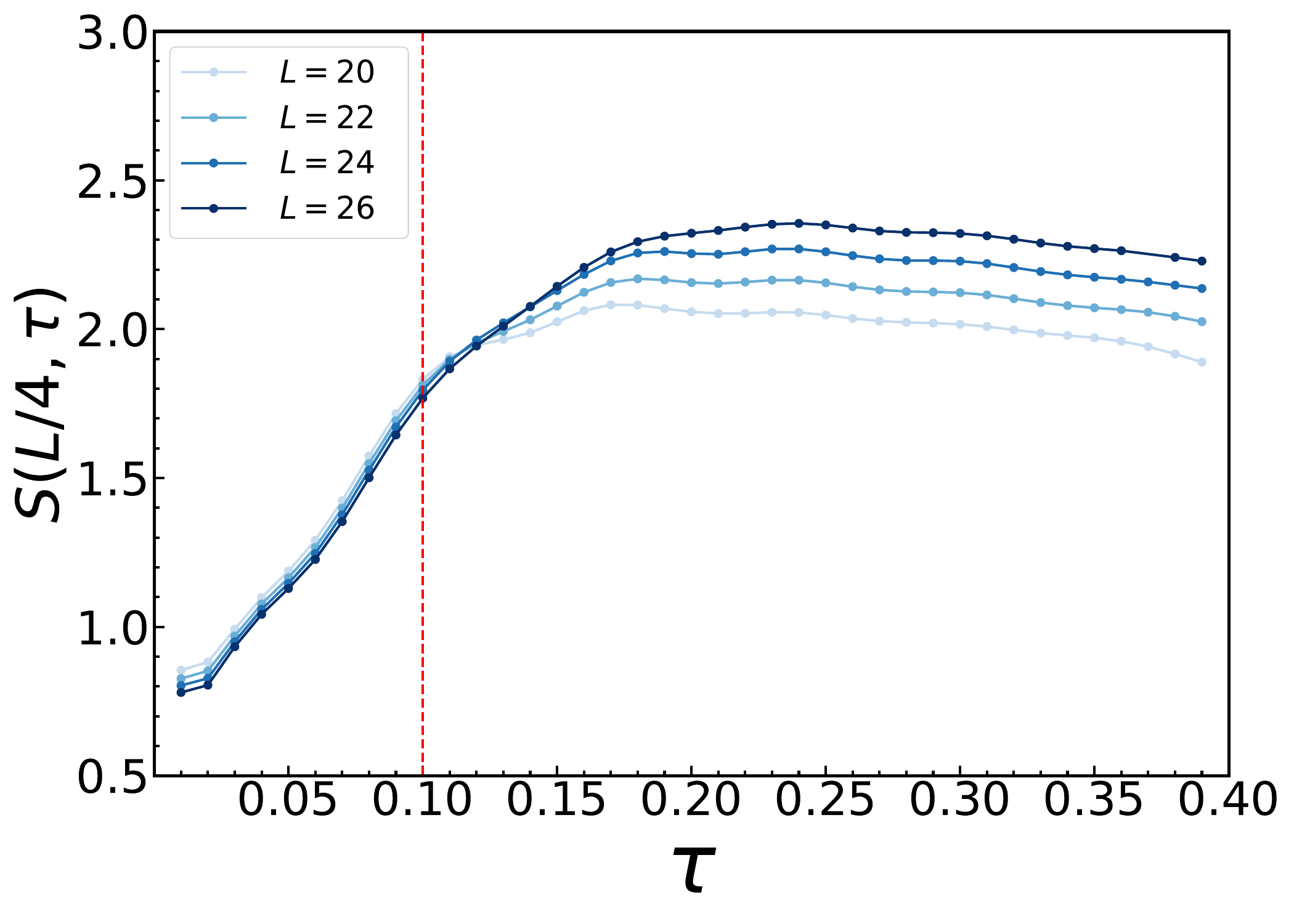}
         \caption{}
         \label{Bi_2}
     \end{subfigure}
      \hfill
     \begin{subfigure}[b]{0.29\textwidth}
         \centering
         \includegraphics[width=\textwidth]{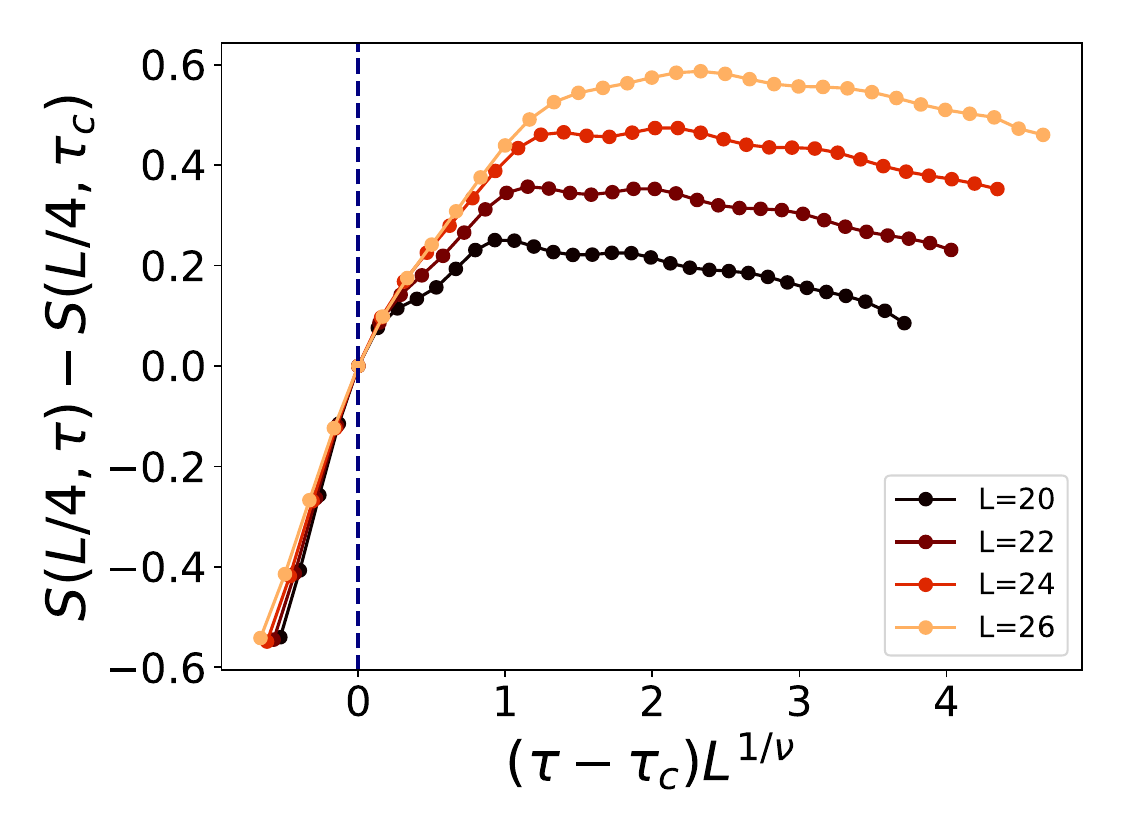}
         \caption{}
         \label{Bi_3}
     \end{subfigure}
      \caption{
      Bipartite entanglement entropy for $L/4$ : rest bipartition ($S(L/4)$) for $h=3/2$ and $n=10$. In panel (a), entanglement entropy  is plotted for different $\tau$ values with respect to the system size $L$. The $\tau$ values increase as the color of the curves becomes deeper. In this context, we have considered values of $\tau$ ranging from $0.06$ to $0.30$, with an increment of $0.02$ between successive $\tau$ values. {Now, among all the $\tau$ values, i.e for $\tau=0.06,0.08,0.10,0.12,0.14,....,0.40$, the curves are almost flat bellow $\tau=0.10$ and above $\tau=0.10$, a almost linear increment of entanglement starts to arise. It is apparent that there is an area law to volume law transition{around $\tau_c=0.20$.} The transition between area-law and volume-law is clear {about the transition point $\tau_c=0.1$.}} It is apparent that there is an area law to volume law transition. In panel (b) we plot $ S(L/4,\tau )$ for different values of $L$
with respect to $\tau$.
       In panel (c) we show the scaling collapse (following Eq. 7 of main text) of the entanglement curves with $\tau_c=0.1$ and exponent {$\nu=0.86$}.
      }
        \label{bipartitec1}
\end{figure}

\begin{figure}
     \centering
     \begin{subfigure}[b]{0.3\textwidth}
         \centering
\includegraphics[width=\textwidth]{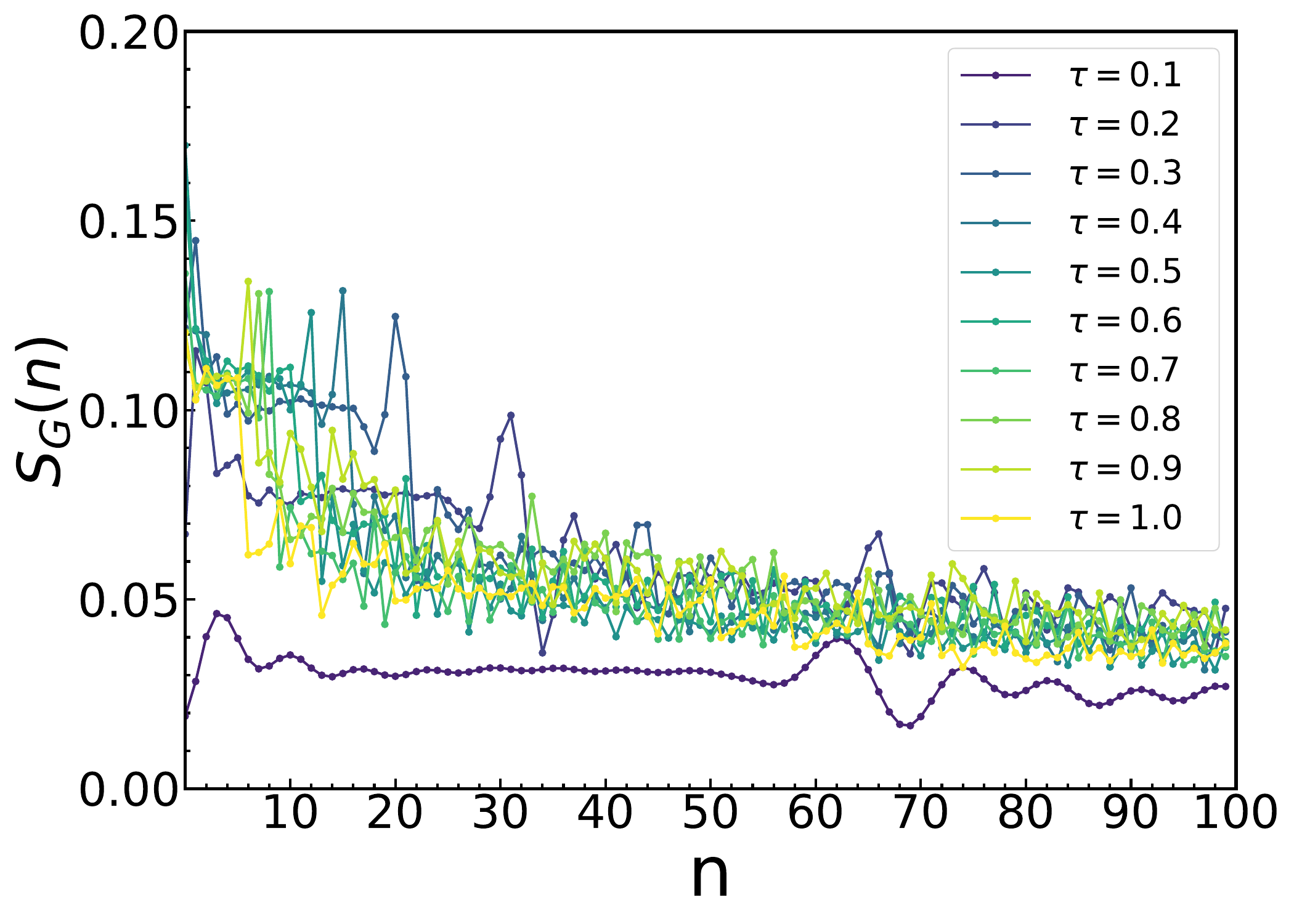}

         \caption{}
         \label{g_1}
     \end{subfigure}
     \hspace{0.9cm}
        \hfill
     \begin{subfigure}[b]{0.3\textwidth}
         \centering
         \includegraphics[width=\textwidth]{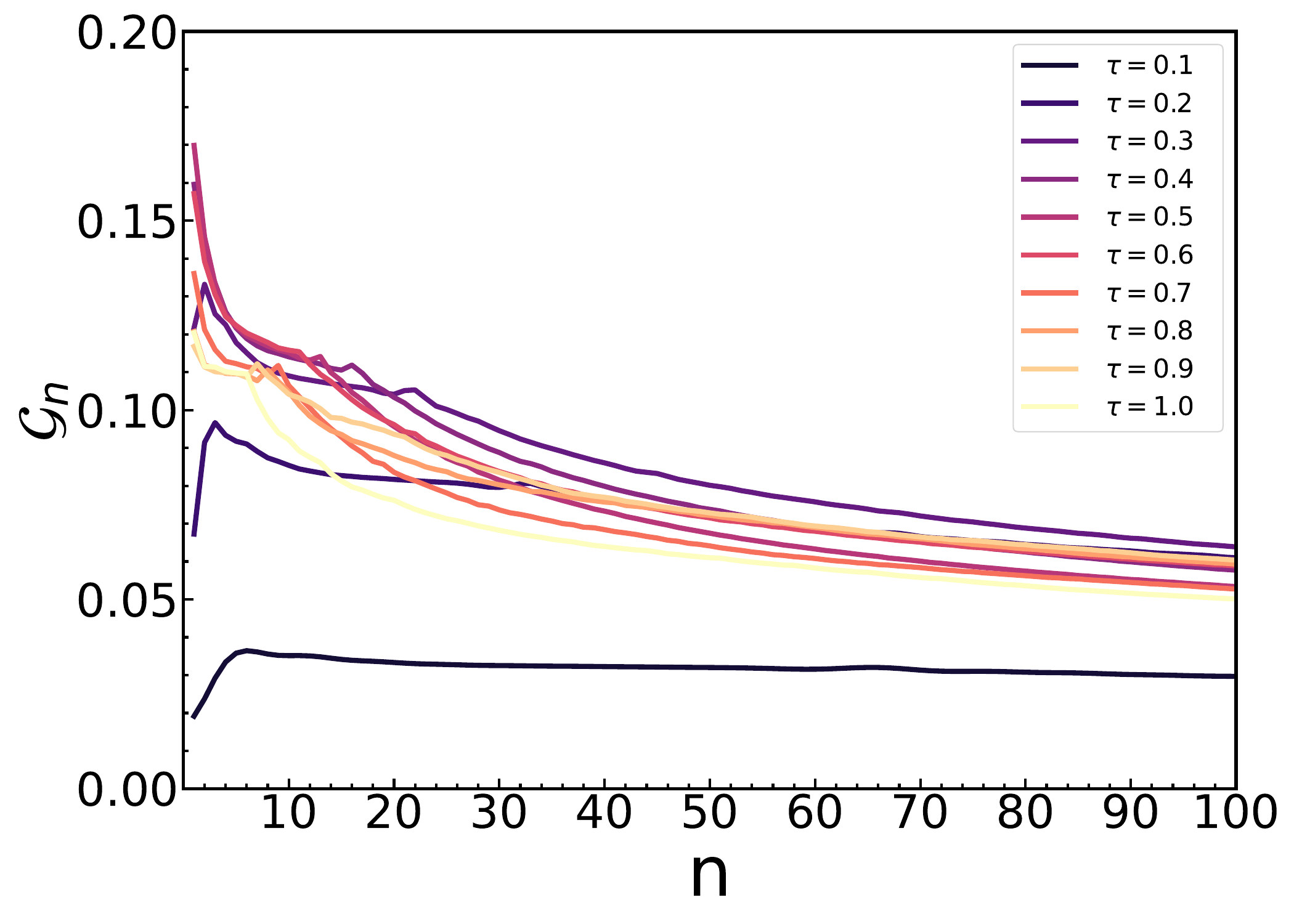}
         \caption{ }
         \label{g_2}
     \end{subfigure}
      \hfill
     \begin{subfigure}[b]{0.3\textwidth}
         \centering
         \includegraphics[width=\textwidth]{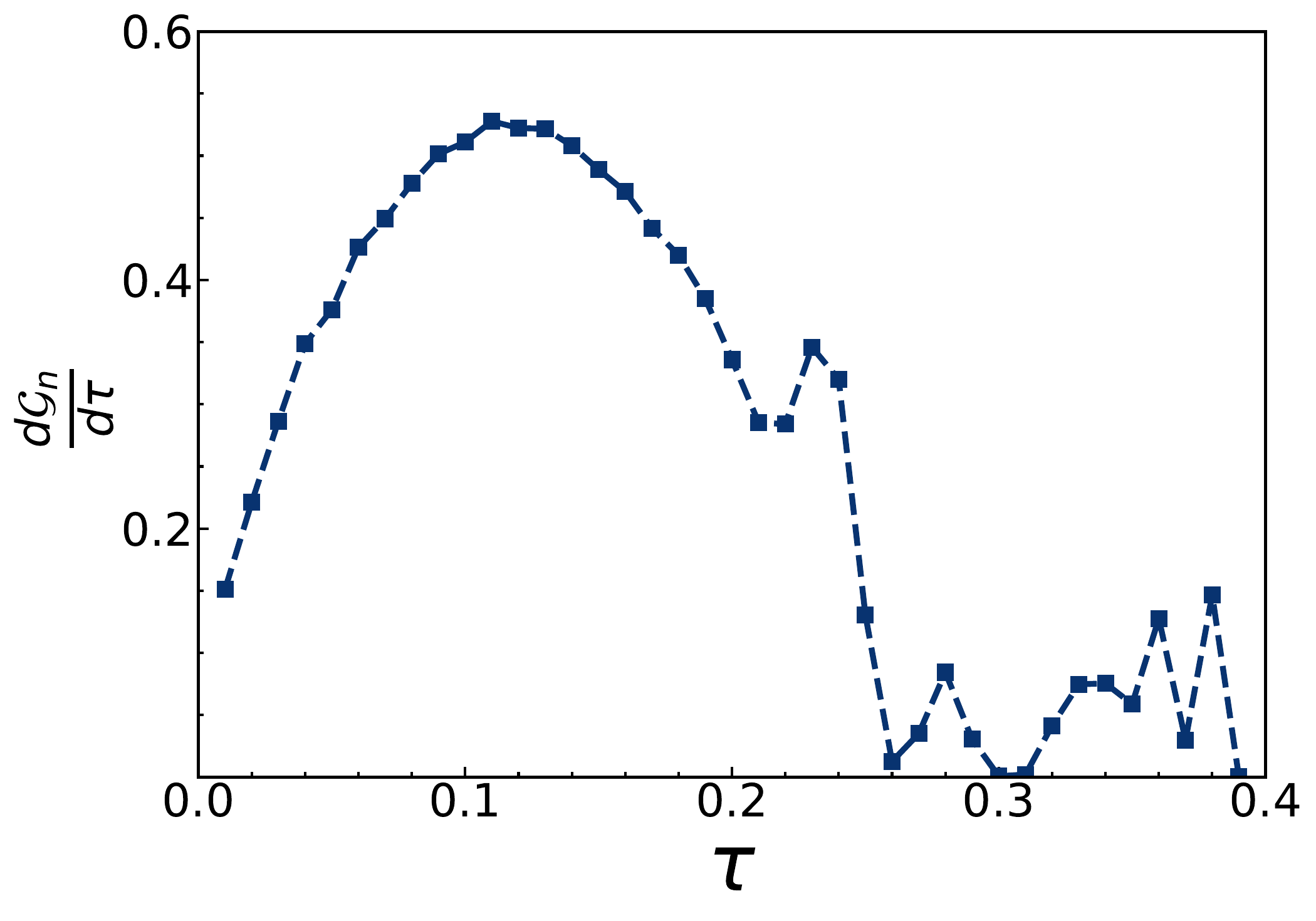}
         \caption{}
         \label{g_3}
     \end{subfigure}
      \caption{ (a)$S_G(n)$ vs $n$ at $h=\frac{3}{2}$ for different $\tau$ values at $L=26$. A gap between the curves corresponding to the $\tau=0.1$ and $\tau=0.2$ values is observed. (b) the cumulative average $\mathcal{G}_n$ of $S_G(n)$ is plotted vs $n$. (c) The plot of $\frac{d\mathcal{G}_n}{d\tau}$ vs $n$ shows the transition point around $\tau_c=0.1$.}
        \label{SM_GGM}
\end{figure}

{{\bf Results for $h=1$:}}\\
The numerical analysis of the survival probability and bipartite entanglement for $h = 1$ is presented in panels (a) and (b) of Fig.~\ref{Rn3d}. In panel (a), we plot $\frac{dH}{d\tau}$ as a function of $\tau$ for system sizes $L = 20, 22, 24,$ and $26$ at field strength $h = 1$. For $L = 26$, the transition point is found to be $\tau_c \approx 0.16$. Panel (b) displays a data collapse of the curves in the plot of $S(L/4,\tau) - S(L/4,\tau_c)$ vs. $(\tau-\tau_c)^{1/\nu}$ at the same field strength $h = 1$. This collapse is obtained by considering system sizes $L = 20, 22, 24,$ and $26$, yielding a critical exponent $\nu = 1.331$ at $\tau_c = 0.16$. These results indicate a clear transition from an area to a volume-law transition about $\tau_c \approx 0.16$.

\begin{figure*}[h!]
		\centering
		\includegraphics[scale=0.21]{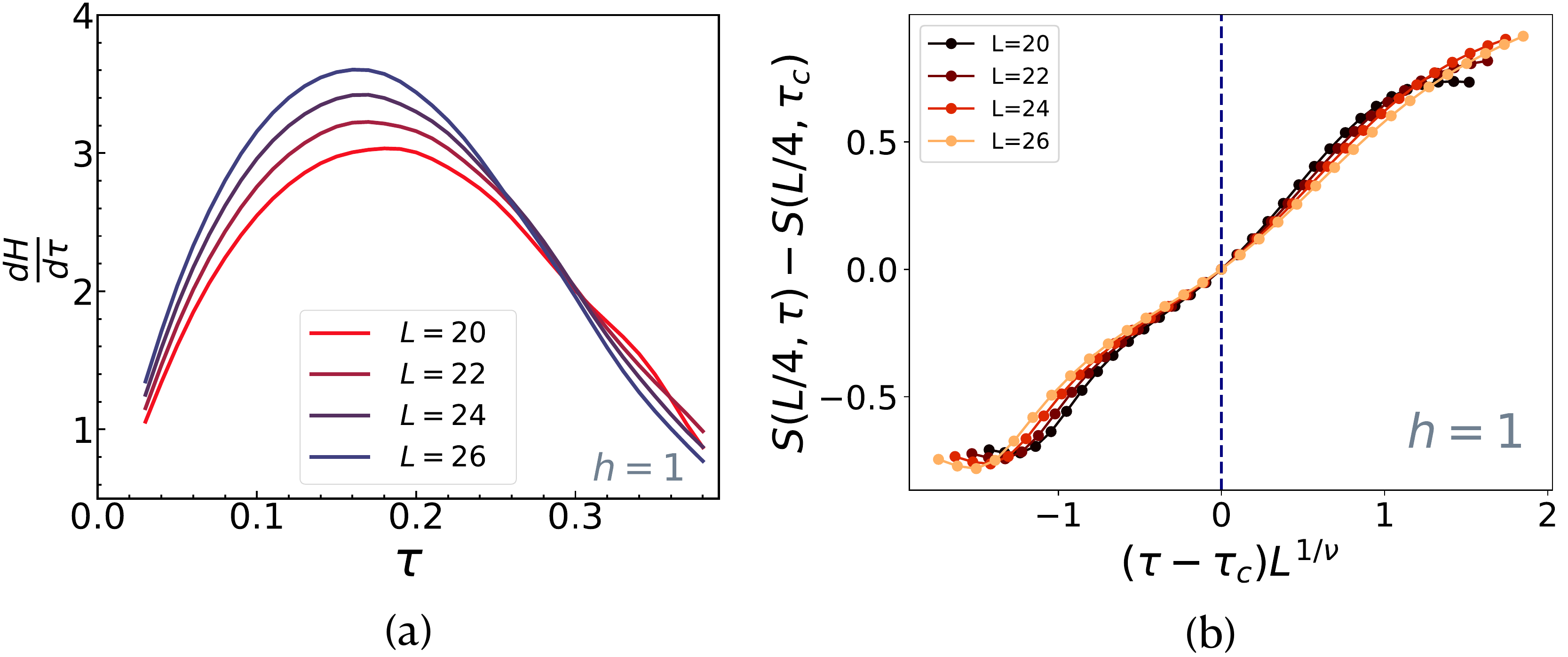}
		\caption{{In panel (a), we have plotted $\frac{dH}{d\tau}$ vs. $\tau$ for $L=20, 22, 24$ and $26$ at the field strength $h=1$. At $h=1$, $\tau_c$ is found to be 0.16 approximately. In panel (b), we show the similar collapse of curves in the $S(L/4,\tau)-S(L/4,\tau_c)$ vs. $(\tau-\tau_c)^{1/\nu}$ plot for field strength $h=1$. This collapse is achieved by considering the system size of the spin chain $L=20, 22, 24$ and $26$. Here the value of the exponent is found to be $\nu=1.331$ at $\tau_c=0.16$.}}
  \label{Rn3d}
	\end{figure*}

\begin{figure}
     \centering
     \begin{subfigure}[b]{0.32\textwidth}
         \centering
         \includegraphics[width=\textwidth]{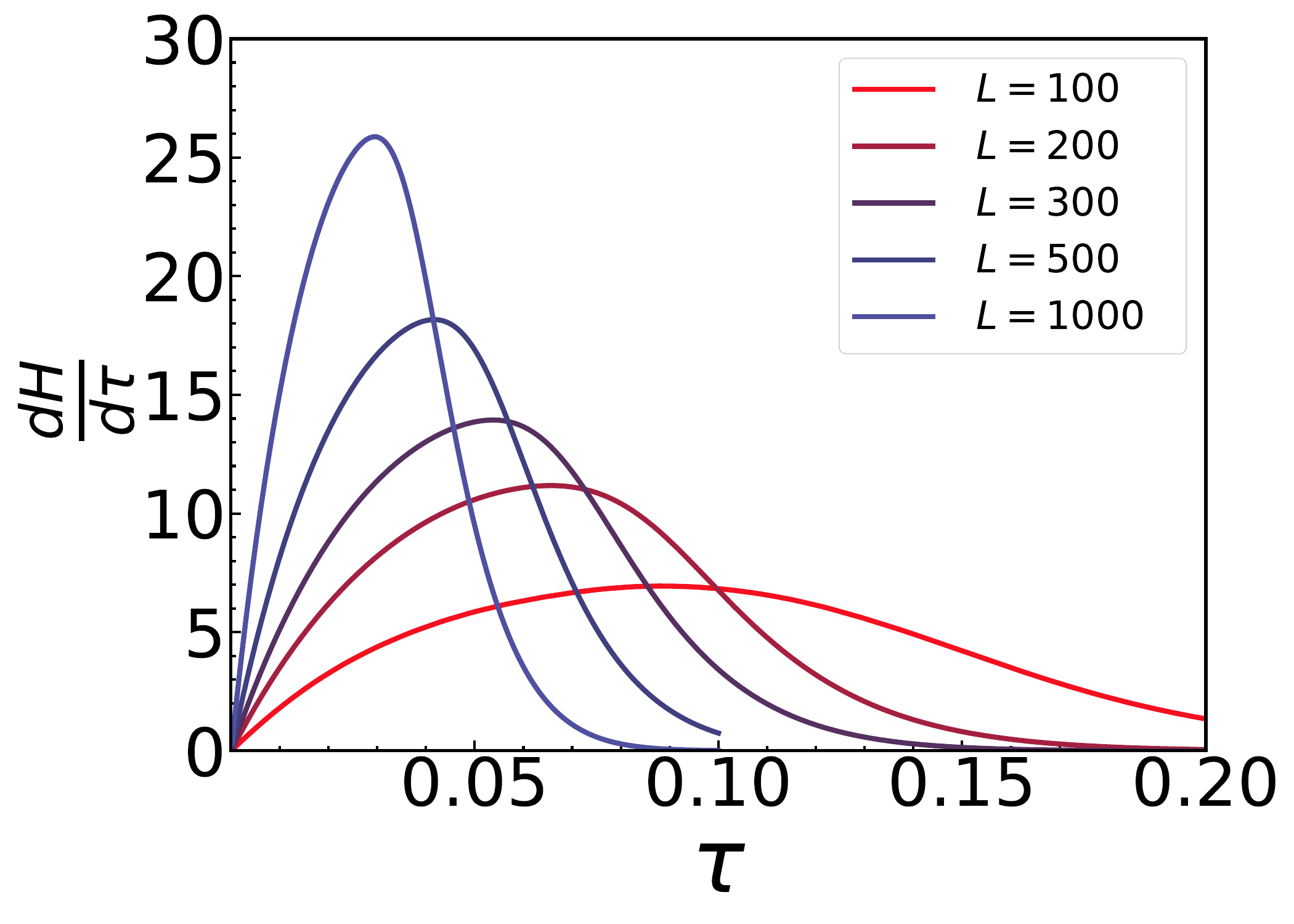}
         \caption{}
         \label{prob_lo_1x}
     \end{subfigure}
     \begin{subfigure}[b]{0.32\textwidth}
         \centering
         \includegraphics[width=\textwidth]{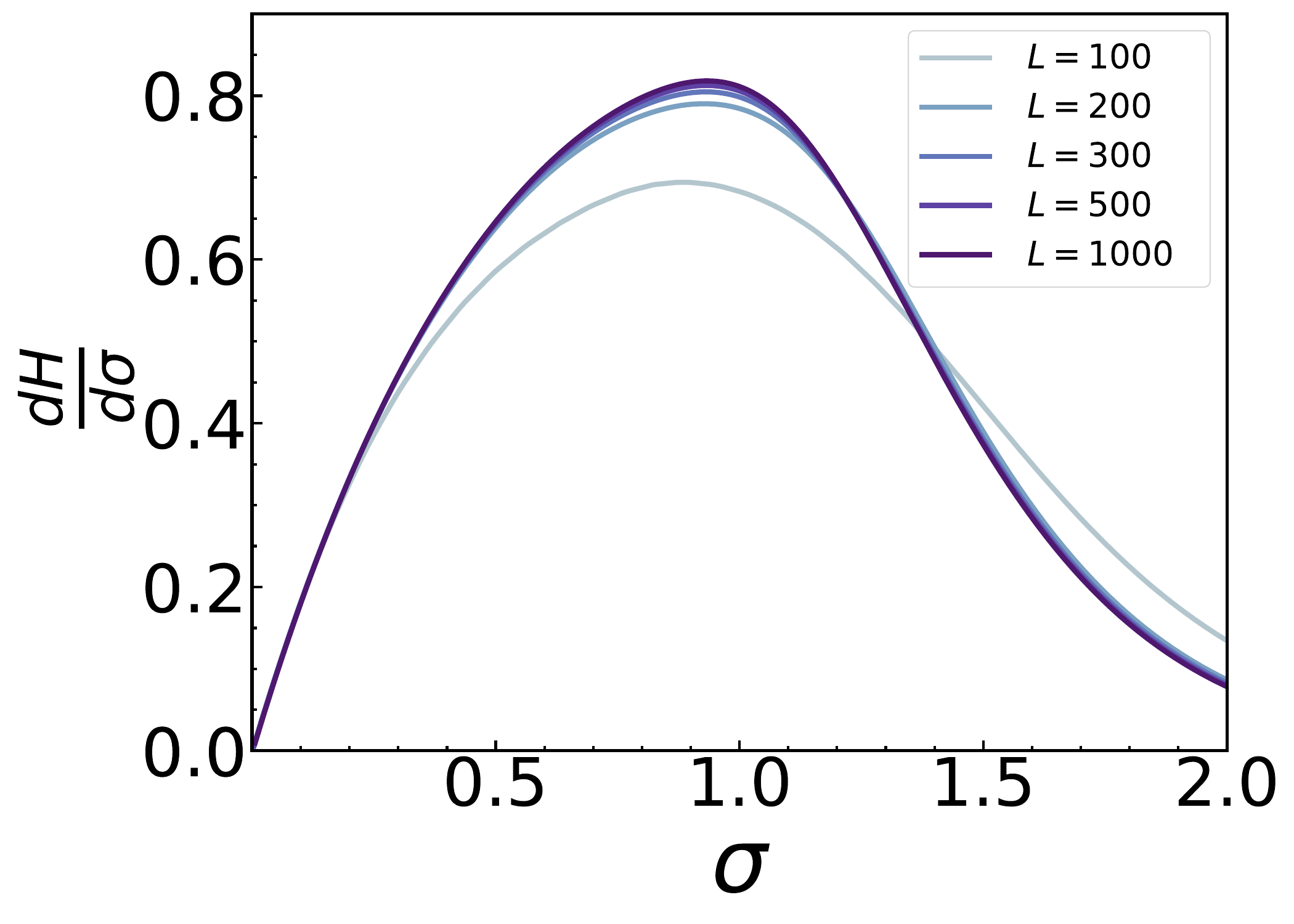}
         \caption{}
         \label{prob_l0_2}
     \end{subfigure}
      \caption{ In panel (a), we have plotted $\frac{dH}{d\tau}$ vs. $\tau$ for $L=100,200,300,500,1000$ at the field strength $h=\frac{3}{2}$. Even at $h=\frac{3}{2}$, $\tau_c$ also moves towards $0$ as $L$ increases. In panel (b), we have plotted $\frac{dH}{d\sigma}$ vs. $\sigma$ at the same system sizes and field strength as FIG.~\ref{prob_lo_1x}, where $\sigma=\tau \sqrt{L}$. All the curves have a peak around $\sigma_c=1$.}
        \label{prob_lof}
\end{figure}

{{\bf Small size scaling of transition point:}} 
\\
{We have already shown in the main manuscript that the transition point varies as $\tau_c = 1/\sqrt{L}$. Here, we have examined how this transition behaves when considering only finite system sizes. In this regard, we have taken system sizes $L = 16, 18, 20, 22, 24, 26,$ and $28$. We have evaluated the scaling function for $h = \frac{1}{2}$ and $h = \frac{3}{2}$ by observing $|\tau_c - \tau_c^{\infty}|$ with system size $L$. Here, $\tau_c^{\infty}$ denotes the transition points at the thermodynamic limit, i.e., $L \rightarrow \infty$. The transition point at the thermodynamic limit for $h=\frac{1}{2}$ and $h=\frac{3}{2}$ is $\tau_c^{\infty}=0.107$ and, $\tau_c^{\infty}=0.103$ respectively. The scaling functions corresponding to $h = \frac{1}{2}$ and $h = \frac{3}{2}$ field strengths have been found to be $|\tau_c - \tau_c^{\infty}| = -0.109 \ln{L} + 0.416$ and $|\tau_c - \tau_c^{\infty}| = -0.081 \ln{L} + 0.317$, respectively. The least squared errors associated with the two fitting curves in Fig.~\ref{prob_lo_1} and Fig.~\ref{prob_l0_2} are given by $0.4405\%$ and $0.443\%$, respectively. Here it is to be noted that for any field strength the analytically achieved scaling function of the transition point, at large size limit, is proportional to $\frac{1}{\sqrt{L}}$, which does not match with the finite size scaling function for both the field strength.}

\begin{figure}
     \centering
     \begin{subfigure}[b]{0.32\textwidth}
         \centering
         \includegraphics[width=\textwidth]{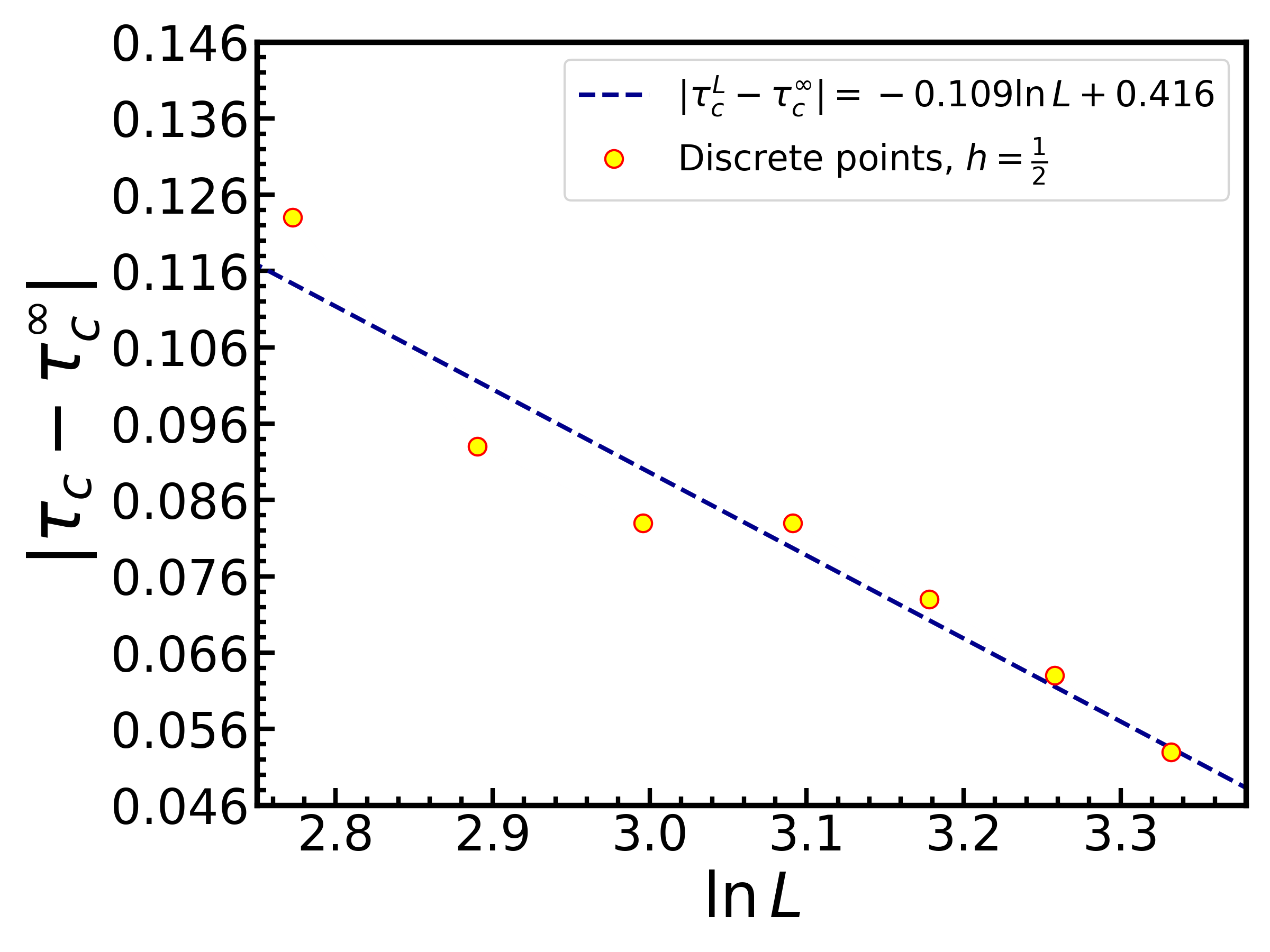}
         \caption{}
         \label{prob_lo_1}
     \end{subfigure}
     \begin{subfigure}[b]{0.32\textwidth}
         \centering
         \includegraphics[width=\textwidth]{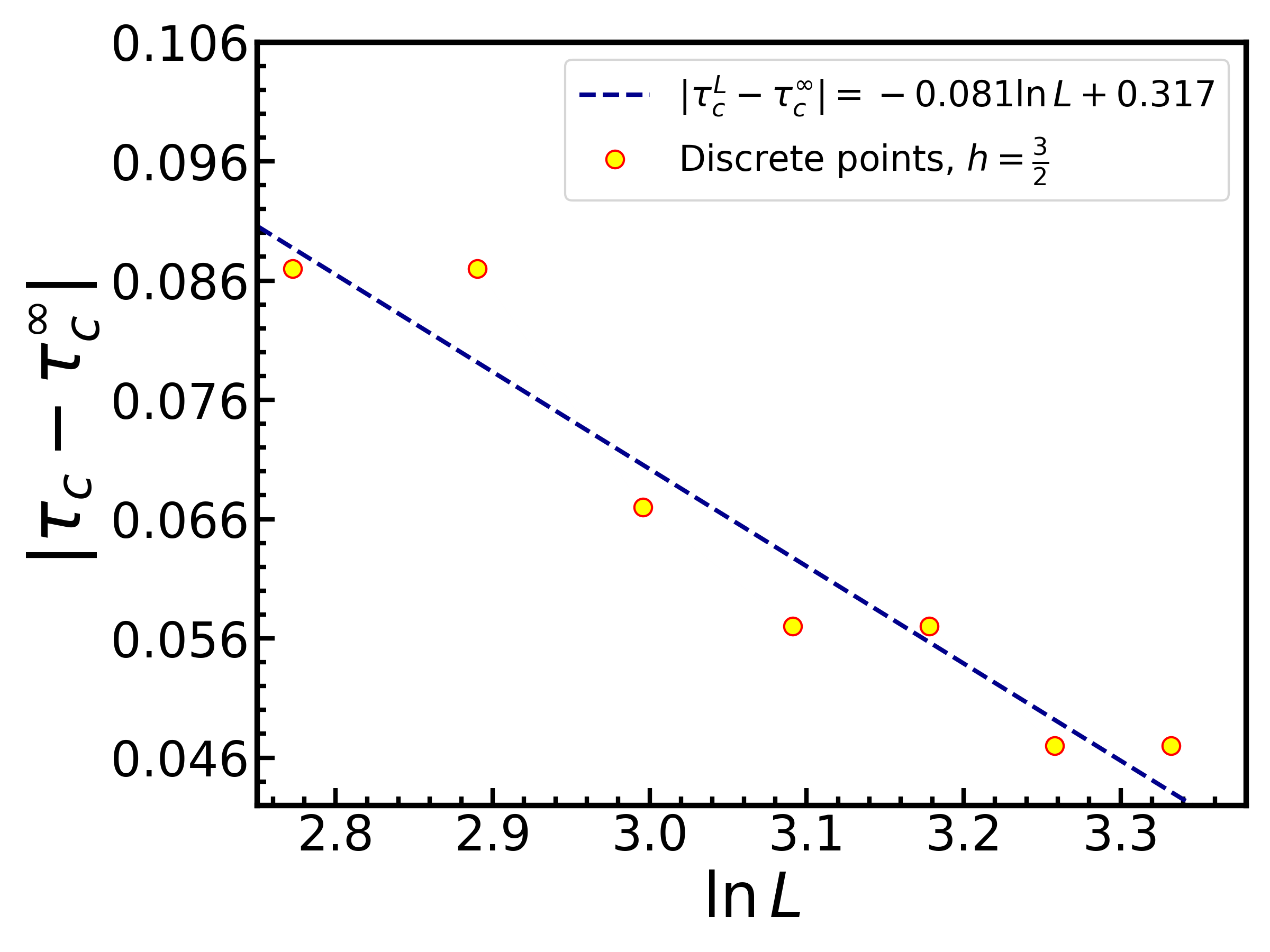}
         \caption{}
         \label{prob_l0_2}
     \end{subfigure}
      \caption{{ Here $\tau^{\infty}_c$ is the transition point at the thermodynamic limit, i.e., $L\rightarrow \infty$. In panel (a), we do finite size scaling of the transition point $\tau_c$ at field strength $h=\frac{1}{2}$. In panel (b), we do the finite size at $h=\frac{3}{2}$ scaling in the same way as in panel (a). In both panels, we use system size $L=16, 18, 20, 22, 24, 26$, and $28$.}}
        \label{prob_lo}
\end{figure}

\begin{figure}
    \centering

    \newcommand{\subfigspace}{0.6cm} 

    \begin{subfigure}[b]{0.3\textwidth}
        \centering
        \includegraphics[width=\textwidth]{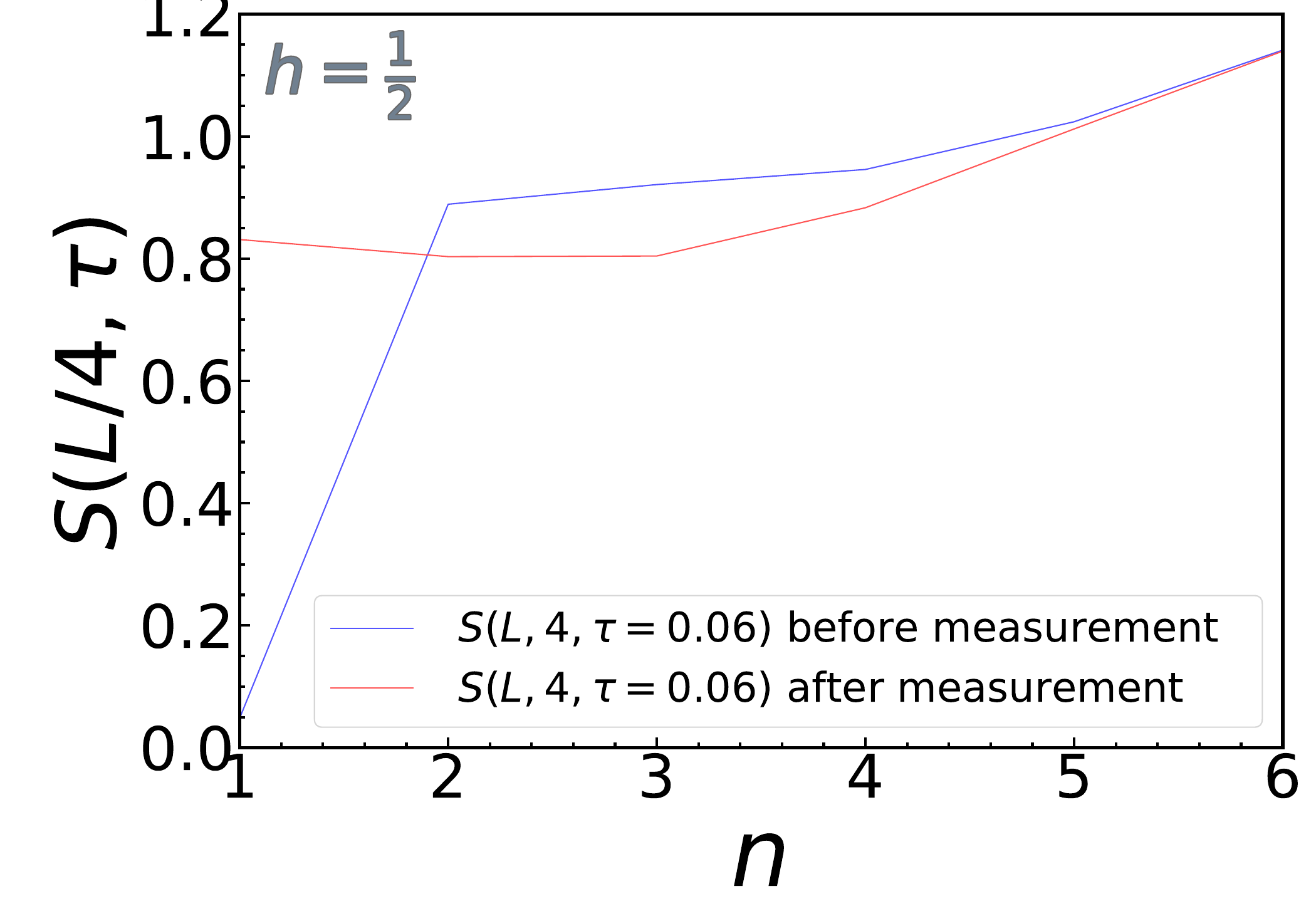}
        \caption{}
        \label{b_1}
    \end{subfigure}
    \hspace{\subfigspace}
    \begin{subfigure}[b]{0.3\textwidth}
        \centering
        \includegraphics[width=\textwidth]{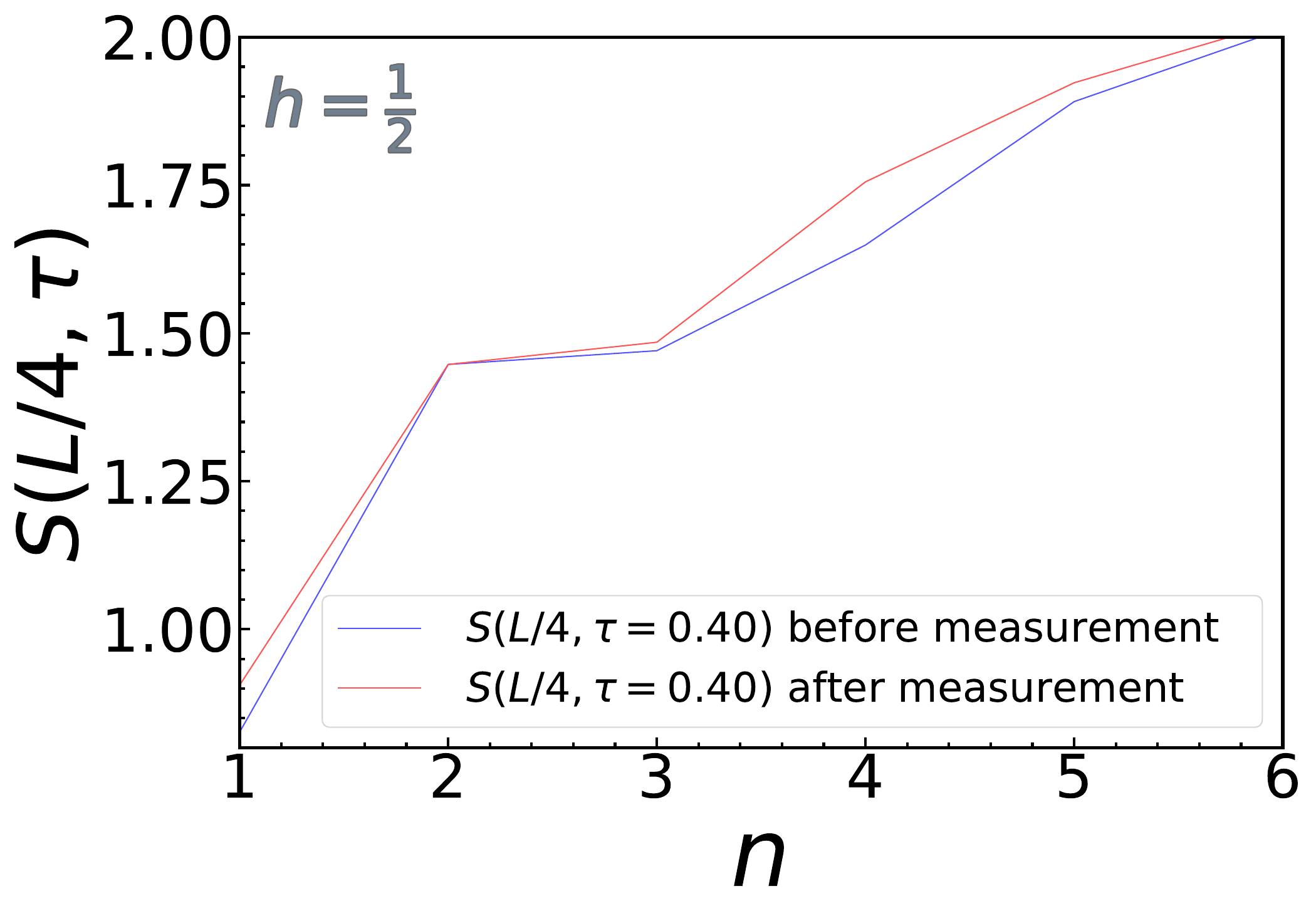}
        \caption{}
        \label{b_2}
    \end{subfigure}
\caption{Bipartite entanglement (of $L/4$: $3L/4$ of the system) just before (blue) and after (orange) the measurement as a function of time-step $n$ for $L=24$ and a) $\tau=0.06$ and b) $\tau=0.4$.}
    \label{prob}
\end{figure}

\begin{figure}
    \centering

    \newcommand{\subfigspace}{0.6cm} 
  \begin{subfigure}[b]{0.3\textwidth}
        \centering
        \includegraphics[width=\textwidth]{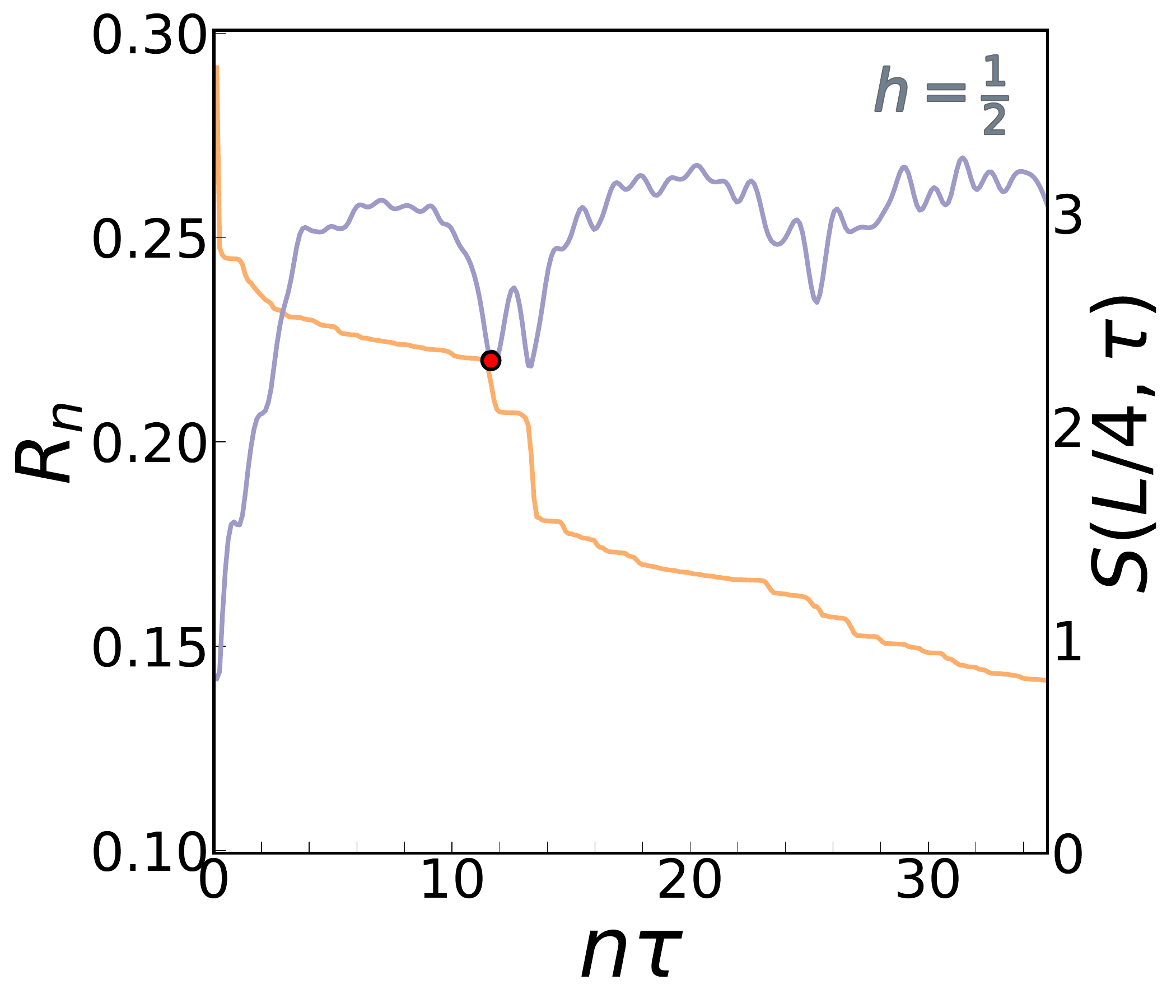}
        \caption{}
        \label{b_1}
    \end{subfigure}
    \hspace{\subfigspace}
    \begin{subfigure}[b]{0.3\textwidth}
        \centering
        \includegraphics[width=\textwidth]{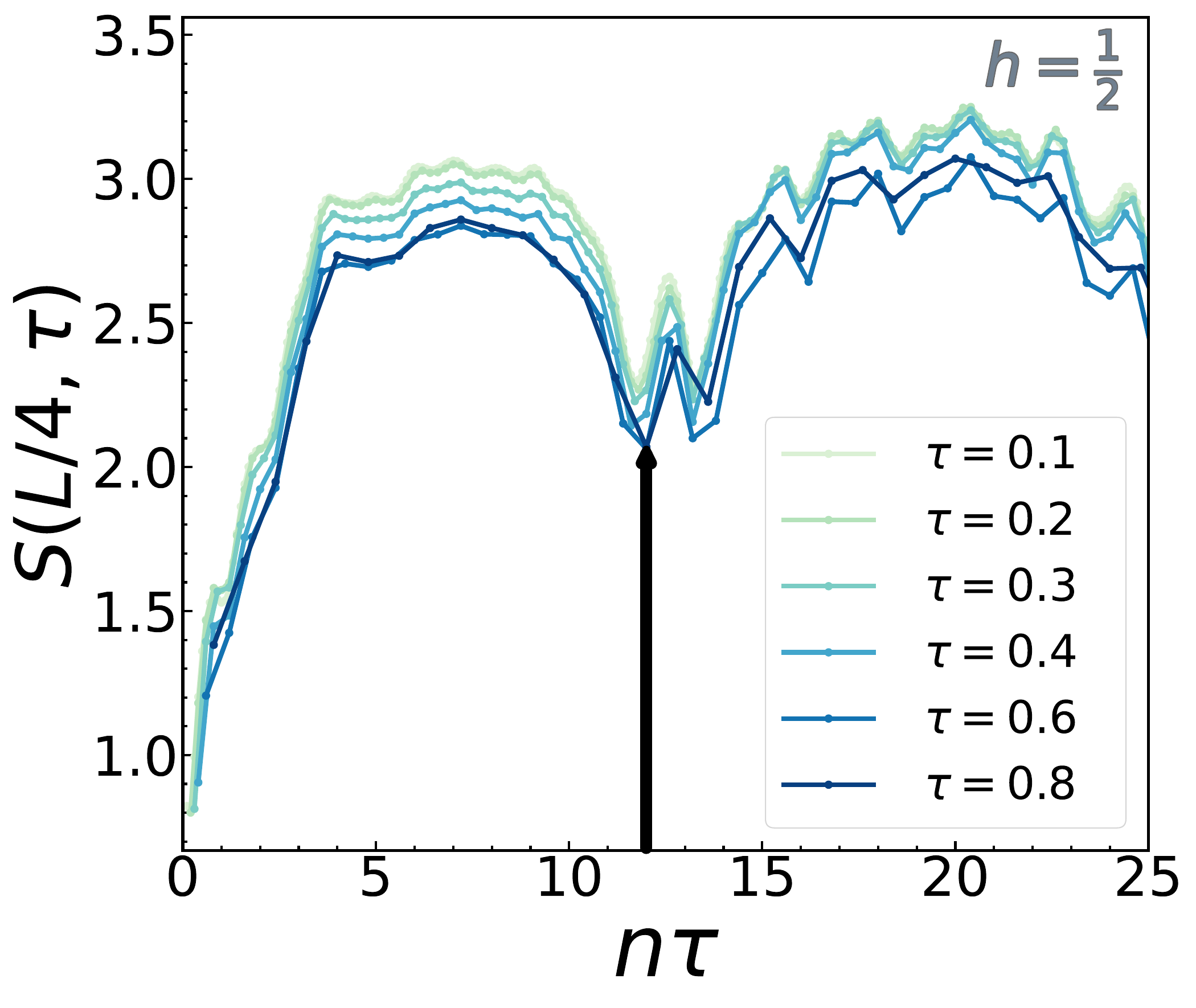}
        \caption{}
        \label{b_2}
    \end{subfigure}
 \caption{{(a) Entanglement (violet) and survival probability (orange)  vs. $n \tau$ for $\tau=0.12$ and $L=24$. This is a typical trend observed for all $\tau$ values investigated. (b) Entanglement vs $n\tau$ is shown for several values of $\tau$.  The first dip, which is actually the end point of the plateau, is marked by a black arrow}.}
    \label{prob_ab}
\end{figure}
{\textbf{Numerical Method of finding the exponent $\nu$:} 
The detailed method of finding the exponent is $\nu$ in our work is as follows. Firstly, to find the data collapse, $n$ numbers of data files ($S$ vs $\tau$) corresponding to each system size $L$ are created. {In our numerical analysis, we have created four data files, i.e $n=4$.} For a given $\tau_c$ and $\nu$, we construct the data sets $S(\tau)-S(\tau_c)$ vs $(\tau-\tau_c)L^{1/\nu}$ for each L. If there is no data for $S$ for the given $\tau_c$, we use linear interpolation to find $S(\tau_c)$. Now we select a window around $\tau_c$, say $\tau_c - \tau_0$ to $\tau_c+\tau_0$, and break it into a number of bins as per our raw data resolution. In each bin we have $S(\tau)-S(\tau_c)$ values for different $L$. We calculate the mean of all the data in a particular bin and calculate the least squared error for each bin and calculate the total least squared error by summing over all the bins. Now we select the final $\tau_c$ and $\nu$ for which the total least squared error is minimum. }

{ \textbf{Difference in the role of measurement with the random circuit model:}} 

{As discussed in the main manuscript, in quantum circuits the transition originates from a competition between the measurement process, which suppresses entanglement, and the unitary evolution, which enhances it. In contrast, in our scenario the transition emerges even though the measurement can either increase or decrease the entanglement, depending on the specific case. This feature is depicted in Fig.~\ref{prob_ab}.}

{\textbf{Dip of entanglement and the end point of the plateau:}}
{Here, we observe an interesting behavior of dip in entanglement in the dynamics and the endpoint of the plateau. In particular, the termination of the plateau region in the $R_n$ vs $n$ plot is accompanied by a sudden dip in entanglement dynamics. This simultaneous occurrence suggests that the existence of the plateau in the $R_n$ vs $n$ plot is closely linked to the behavior of the entanglement entropy.
This feature is pictorially shown in Fig.~\ref{prob_ab}.}

\twocolumngrid
\bibliography{MIPT}

\end{document}